\documentclass{aa}
\usepackage{graphicx}
\usepackage{xcolor}
\usepackage{natbib}
\usepackage{float}
\usepackage{xspace}
\usepackage{mhchem}
\usepackage{textcomp}
\newcommand{\change}[1]{#1}

\usepackage{savesym}
\usepackage{amsmath}
\savesymbol{iint}
\usepackage{txfonts}
\restoresymbol{TXF}{iint}

\begin{document}

\title{Isotope selective photodissociation of \ce{N2} by the interstellar radiation field and cosmic rays}
\author{Alan~N.~Heays\inst{1} \and Ruud~Visser\inst{2} \and Roland~Gredel\inst{3} \and Wim~Ubachs\inst{4} \and Brenton~R.~Lewis\inst{5} \and Stephen~T.~Gibson\inst{5} \and Ewine~F.~van~Dishoeck\inst{1,6} }
\institute{
  $^1$ Leiden Observatory, Leiden University, P.O. Box 9513, 2300 RA Leiden, The Netherlands\\
  $^2$ Department of Astronomy, University of Michigan, 500 Church Street, Ann Arbor, MI 48109-1042, USA \\
  $^3$ Max-Planck-Institut f\"ur Astronomie (MPIA), Königstuhl 17, D-69117 Heidelberg, Germany \\
  $^4$ Department of Physics and Astronomy, LaserLaB, VU University, de Boelelaan 1081, 1081 HV Amsterdam, The Netherlands\\
  $^5$ Research School of Physics and Engineering, The Australian National University, Canberra, ACT 0200, Australia \\
  $^6$ Max-Planck Institut f\"ur Extraterrestrische Physik (MPE), Giessenbachstra\ss{}e.\ 1, 85748 Garching, Germany \\
  \email{heays@strw.leidenuniv.nl}}
\date{\textbf{Draft -- \today}}
\titlerunning{Isotope selective photodissociation of \ce{N2}}
\authorrunning{Heays et al.}

\abstract
{Photodissociation of \ce{^{14}N2} and  \ce{^{14}N ^{15}N} occurs in interstellar clouds, circumstellar envelopes, protoplanetary discs, and other environments due to ultraviolet radiation originating from stellar sources and the presence of cosmic rays. This source of N atoms initiates the formation of more complex N-bearing species and may influence their isotopic composition.}
{To study the photodissociation rates of \ce{^{14}N ^{15}N} by ultraviolet continuum radiation and both isotopologues in a field of cosmic ray induced photons. To determine the effect of these on the isotopic composition of more complex molecules.}
{High-resolution theoretical photodissociation cross sections of N$_2$ are used from an accurate and comprehensive quantum-mechanical model of the molecule based on laboratory experiments, as presented for \ce{^{14}N2} Li et al. 2013.
A similarly high-resolution spectrum of H$_2$ emission following interactions with cosmic rays has been constructed.
The spectroscopic data are used to calculate photodissociation rates which are then input into isotopically differentiated chemical models, describing an interstellar cloud and a protoplanetary disc.  }
{The photodissociation rate of \ce{^{14}N ^{15}N} in a Draine field assuming 30\,K excitation is $1.73\times 10^{-10}\,\mathrm{s}^{-1}$, within 4\% of the rate for \ce{^{14}N2}, and the rate due to cosmic ray induced photons \change{assuming an \ce{H2} ionisation rate of $\zeta=10^{-16}$\,s$^{-1}$} is about $10^{-15}\,\mathrm{s}^{-1}$, with up to a factor of 10 difference between isotopologues.
Shielding functions for \ce{{}^{14}N^{15}N} by \ce{^{14}N2}, \ce{H2}, and H are presented.
\change{Incorporating these into an interstellar cloud model, an enhancement} of the atomic $\mathrm{{}^{15}N/{}^{14}N}$ ratio over the elemental value is obtained due to the self-shielding of external radiation at an extinction of about 1.5\,mag.
This effect is larger where assumed grain growth has reduced the opacity of dust to ultraviolet radiation.
The transfer of photolytic isotopic fractionation of N and N$_2$ to other molecules is demonstrated to be significant \change{in a protoplanetary disc model with grain growth}, and is species dependent with \ce{^{15}N} enhancement approaching a factor of 10 for HCN.
\change{The cosmic ray induced dissociation of CO is revisited employing a more recent photodissociation cross section, leading to a rate that is $\sim$40\% lower than previously calculated.}
}
{}
 \maketitle

\section{Introduction}
\label{sec:intro}

Nitrogen is the fifth most abundance element in the universe and comprises two isotopes with relative populations ${}^{14}\mathrm{N}/{}^{15}\mathrm{N}\simeq 400$.
The dominant reservoir of elemental N is in the form of atoms and N$_2$, which are essentially unobservable outside of the solar system, the former because its ionic form, N$^+$, is readily removed by charge exchange with neutral hydrogen with the latter lacking a rotational emission spectrum because it is homonuclear.
The observation of electronic transitions of N$_2$ is limited to technically-challenging ultraviolet (UV) wavelengths shorter than 1000\,\AA{}.

There have been numerous detections of N-derived molecules such as \ce{N2H+}, $\mathrm{CN}$, $\mathrm{HCN}$, and \ce{NH3}  in interstellar space, protostellar envelopes and discs, and extra-galactically, \citep[e.g.,][]{van_zadelhoff2001,bergin2002,gerin2009,oberg2010,tobin2012}, as well as for their isotopologues \citep[e.g.,][]{wannier1991,gerin2009,adande2012,hily-blant2013,bizzocchi2013,daniel2013}.
The balance of chemically active N and highly stable N$_2$ is an important parameter in the chemistry of these more complex amines and nitriles.
The primary driver of this balance in environments exposed to an ultraviolet flux is photodissociation of N$_2$.

The value of $\mathrm{{}^{14}N}/\mathrm{^{15}N}$ in observable species constitutes an astronomical puzzle.
The solar elemental ratio, 440 \citep{marty2010}, is considered representative of the local interstellar medium, but differs from the terrestrial value, 270; in the atmospheric of Titan, 180 \citep{niemann_etal2005}; and in interplanetary dust, meteorites and comets, \change{68--330 \citep{floss2006,jehin2009,aleon2010,mumma2011}}.
Observational surveys have been made of galactic star-forming regions \citep{adande2012} and dark clouds \citep{hily-blant2013} and also point to a range of values, as well as variations by source and the molecular species observed.
Some variation of ${}^{14}$N/${}^{15}$N is clearly due to the details of local nuclear synthesis, with an increasing excess of ${}^{15}$N towards the galactic centre \citep{wilson1999,adande2012}.
Another proposed explanation for the observed ratios of $\mathrm{C^{14}N}/\mathrm{C^{15}N}$, $\mathrm{HC^{14}N}/\mathrm{HC^{15}N}$ and $\mathrm{{}^{14}NH_3}/\mathrm{{}^{15}NH_3}$ is low-temperature isotopic-exchange chemistry.
Models considering this mechanism \change{\citep[e.g.,][]{rodgers2008,wirstrom2012,le_gal2013}} have been successful in introducing strongly non-elemental abundance ratios under specific conditions.
The critical exchange reactions, e.g.,
\begin{equation*}
  \rm
  HC^{14}NH^+ + {}^{15}N \leftrightharpoons HC^{15}NH^+ + {}^{14}N + energy,
\end{equation*}
rely on the lower zero-point energy of the heavier molecular species and are exothermic by $\leq30$\,K \citep{terzieva2000}, limiting their significance to cold environments.
This is in contrast to similar reactions affecting the H/D abundance in molecules which may proceed at temperatures as high as 170\,K.

An alternative route to the mass fractionation of molecules is isotope selective photodissociation. 
For the case of $\mathrm{N_2}$ dissociation, self-shielding has been shown to drive the $\mathrm{HC^{14}N}/\mathrm{HC^{15}N}$ ratio in the atmosphere of Titan \citep{liang_etal2007}.
Additionally, there is strong observational evidence that the fractionation of protostellar oxygen isotopes is assisted by the self-shielding of CO, which has very similar molecular structure to N$_2$ \citep{sheffer2002,lyons_young2005,smith2009}.
The investigation of isotope selective N$_2$ photodissociation in interstellar clouds and protoplanetary environments is then highly warranted.

The importance of photodissociation due to the interstellar radiation field (ISRF) or an ultraviolet emitting star will decrease as the radiation is attenuated below the surface of an interstellar cloud, circumstellar envelope, or protoplanetary disc.
\citet{li2013} made a detailed assessment of the dissociation rate of \ce{^{14}N2} and the shielding of ultraviolet radiation by relevant materials (H$_2$, H, \ce{^{14}N2} itself, and dust grains).
This work considered in detail the sharply structured absorption spectrum of N$_2$ and shielding species.
This was made possible by means of an accurate and comprehensive quantum-mechanical model of N$_2$ photoabsorption and dissociation, which was informed by an extensive history of N$_2$ spectroscopy in the laboratory \citep[e.g.,][]{ajello_etal1989,helm_etal1993,sprengers_etal2003,sprengers_etal2004b,sprengers_etal2005b,stark_etal2008,lewis_etal2008a,heays_etal2009,heays2011b} and theoretically \citep[e.g.,][]{spelsberg_meyer2001,lewis_etal2005a,lewis_etal2005b,haverd_etal2005,lewis_etal2008c,lewis_etal2008b,ndome_etal2008}. This model, detailed in Sec.~\ref{sec:CSE model}, explicitly accounts for the nuclear masses and so is equally applicable to the case of \ce{^{14}N ^{15}N}. In Sec.~\ref{sec:pd rates} of this paper, we describe new calculations of \ce{^{14}N ^{15}N} photodissociation rates in interstellar and blackbody radiation fields.

The interior regions of interstellar clouds and protoplanetary discs may be completely shielded from external sources of ultraviolet photons.
However, collisions between high-energy cosmic rays and \ce{H2} produce a weak local UV field that could be important for photodissociation.
The cosmic ray induced photoionisation and photodissociation of molecular and atomic species is well established \citep{gredel1987,gredel1989,dalgarno2006,padovani2009}.
In Sec.~\ref{sec:CR} we combine our model of N$_2$ photoabsorption with a similarly thorough theoretical treatment of the H$_2$ line emission associated with cosmic rays.
This is, to study in detail for the first time the photodissociation of \ce{^{14}N2} and \ce{^{14}N ^{15}N} in cloud and disc interiors. 
\change{We also revisit calculations of the CO dissociation rate due to cosmic rays \citep{gredel1987} taking advantage of an updated photodissociation cross section.}

In Sec.~\ref{sec:photochemical models} of this paper we present the results of chemical models of an interstellar cloud and protoplanetary disc. 
These include linked chemical networks for species containing ${}^{14}$N and ${}^{15}$N and the photodissociation rates of \citet{li2013} and Secs.~\ref{sec:pd rates} and \ref{sec:CR}.
Using these models we investigate the consequences of isotope selective photodissociation on the abundances of atomic and molecular nitrogen and other N-bearing species observed in clouds and discs.

\section{The model of photoabsorption and photodissociation}
\label{sec:CSE model}

The photodissociation of N$_2$ first requires excitation into bound states that then rapidly predissociate.
Thus, the photoabsorption spectrum contains well separated electronic-rovibrational lines and must be treated at high resolution ($\mathord{\sim}\!10^{-3}$\,\AA) but over a comparatively large wavelength range (912--1000\,\AA).
Actually, \ce{N2} absorbs strongly at considerably shorter wavelengths but the present application is limited to longwards of the atomic H ionisation threshold at 912\,\AA{}.
Here, a broadband and detailed spectrum is calculated by means of a quantum mechanical model which solves a coupled Schr\"odinger equation (CSE) for the vibrational wavefunction of the molecule.
Properties of the electronic wavefunction are described by potential-energy curves and state coupling parameters optimised with respect to a large database of laboratory line positions, oscillator strengths, and predissociation rates.
These empirical parts of the CSE model formulation are independent of molecular mass, and so are unchanged by isotopic substitution of one or both nuclei.
Then, the explicit calculation of vibrational wavefunctions models the quantum mixing of electronic states and its dependence on the nuclear masses.
A more detailed description of the method, the important electronic states and potential-energy curves used in the model is given in our previous publication \citep{li2013}, and the chemical physics literature \citep{lewis_etal2005a,lewis_etal2005b,haverd_etal2005,lewis_etal2008b,heays2011} with specific discussion of \ce{^{14}N ^{15}N} given in \citet{vieitez_etal2008a} and \citet{heays2011b}.

\begin{figure}
\centering
\includegraphics[width=0.9\hsize]{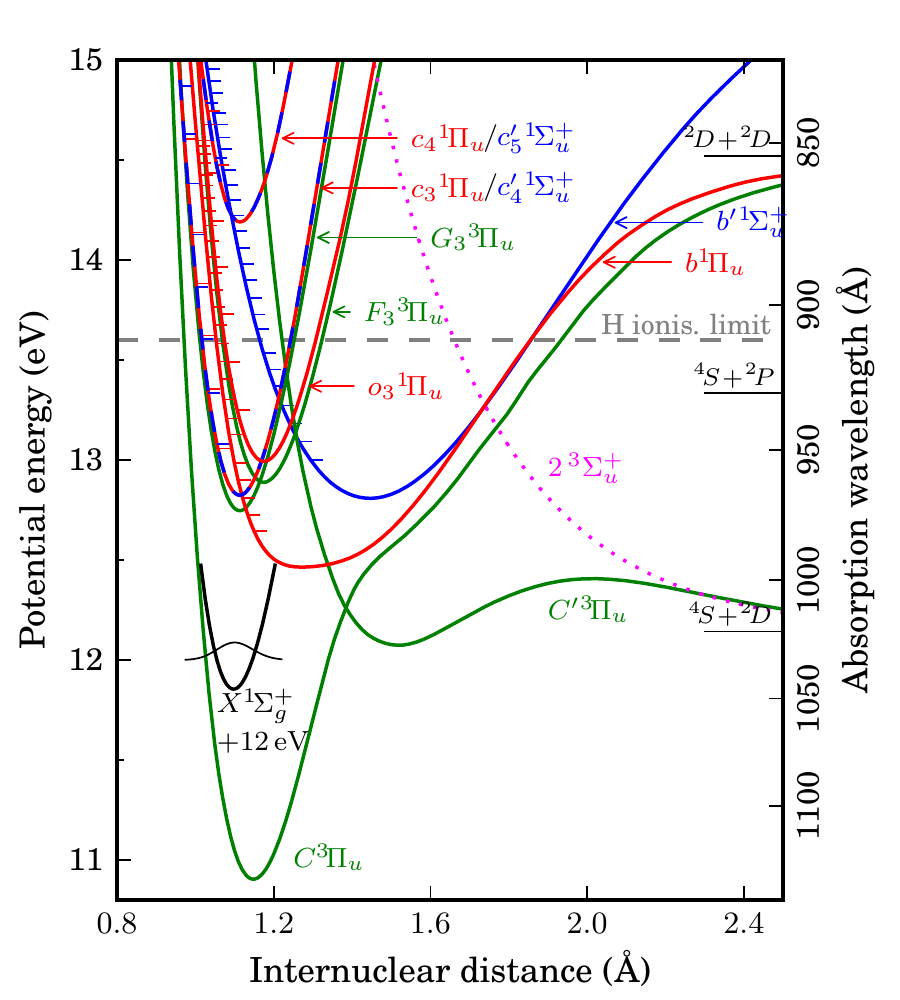}
\caption{Diabatic-basis potential-energy curves of N$_2$ excited states relevant to ultraviolet photoabsorption and photodissociation. The energy scale is referenced to $v=0$, $J=0$ of the $X\,{^1\Sigma_g^+}$ ground state and is related to the wavelength scale assuming transitions from this fundamental level. Also indicated are the excitation states of atomic products arising from dissociation and the absorption limit imposed by the ionisation of atomic H. The potential-energy curves of the $c_n{}^1\Pi_u$ and $c'_{n+1}{}^1\Sigma^+_u$ states are indistinguishable at the plotted scale. The \ce{N2} ionisation potential energy occurs at 15.4\,eV in this figure.}
\label{fig:N2potentials}
\end{figure}

Electric-dipole-allowed photoabsorption from the ground state occurs for wavelengths shorter than 1000\,\AA{} and accesses two valence states and several Rydberg series of $^1\Pi_u$ and $^1\Sigma_u^+$ symmetry.
Potential-energy curves depicting these states are shown in Fig.~\ref{fig:N2potentials} with the lowest-energy photoabsorbing state, $b\,{}^1\Pi_u$, occurring at 12.6\,eV.
There is strong electronic coupling between valence and Rydberg states of the same symmetry which leads to a very perturbed spectrum, including constructive and destructive interference in the absorption strength and predissociation rates of vibrational bands \citep{dressler1969,spelsberg_meyer2001,lewis_etal2005a}.
Additional rotational interactions between $^1\Pi_u$ and $^1\Sigma_u^+$ states lead to strong rotational dependence of line strengths and predissociation rates (or linewidths) \citep{stark_etal2005,heays2011b}.

\begin{figure}
\centering
\includegraphics{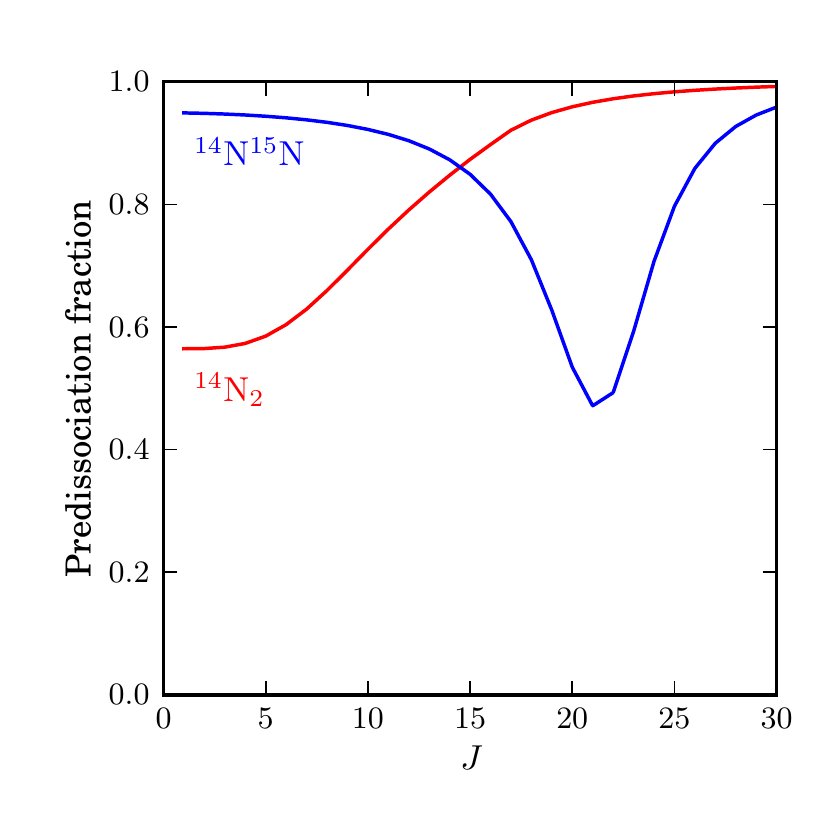}
\caption{An example of isotopically-dependent predissociation. The predissociation fraction\change{, $\eta^\text{pre}$,} of individual rotational levels, $J$, of $b\,{}^1\Pi_u(v=1)$ calculated by the CSE model for \ce{^{14}N2} and \ce{{}^{14}N^{15}N}. }
\label{fig:b01 predissoc frac}
\end{figure}

Predissociation occurs through spin-orbit coupling of ${}^1\Pi_u$ states with a number of states with ${}^3\Pi_u$ symmetry, including an unbound level, and one state of ${}^3\Sigma_u^+$ symmetry.
These are shown in Fig.~\ref{fig:N2potentials} and lead to dissociation into N atoms in the ${}^4S$, ${}^2D$, or ${}^2P$ states.
The spin-selection rules forbid the appearance \ce{N2} triplet states in optical spectra although many have been observed in the laboratory regardless, due to their coupling with ${}^1\Pi_u$  levels \citep{sprengers_etal2005b,lewis_etal2008a}.
The majority of excited vibrational states predissociate after an absorption event with near certainty, that is, they have a predissociation fraction of $\mathord{\sim}1$.
\change{For those states that decay substantially by emission, a reduction of the CSE-calculated photoabsorption cross section, $\sigma^\text{abs}$, has been made by their less-than-unity predissociation fractions, $\eta^\text{pre}$, leading to a photodissociation cross section $\sigma^\text{pd}$, so that
\begin{equation}
  \sigma^\text{pd} = \eta^\text{pre} \sigma^\text{abs}.
\end{equation}
}
This provides an example of the utility of the physically-based CSE model.
That is, the $b\,{}^1\Pi_u(v=1)$ level of \ce{^{14}N2} is known to predissociate significantly less rapidly than for \ce{^{14}N ^{15}N} \citep{sprengers_etal2004b,sprengers_etal2005,wu2011} because of mass-change-induced shifts of the interacting ${}^1\Pi_u$ and ${}^3\Pi_u$ levels.
These shifts are evaluated explicitly by the CSE model and calculated predissociation fractions for $b\,{}^1\Pi_u(v=1)$ are shown in Fig.~\ref{fig:b01 predissoc frac} \change{for rotational levels with angular momentum quantum number, $J\leq 30$.}
There is a strong dependence on rotational angular momentum, $J$, for both isotopologues.
At the low temperatures of interstellar space only rotational levels with $J<10$ are accessible.
Then, photons absorbed by \ce{^{14}N ^{15}N} and leading to excitation of the $b\,{}^1\Pi_u(v=1)$ level are nearly twice as likely to lead to a dissociation event than for \ce{^{14}N2}.

\begin{figure*}
  \centering
  \includegraphics[width=\textwidth]{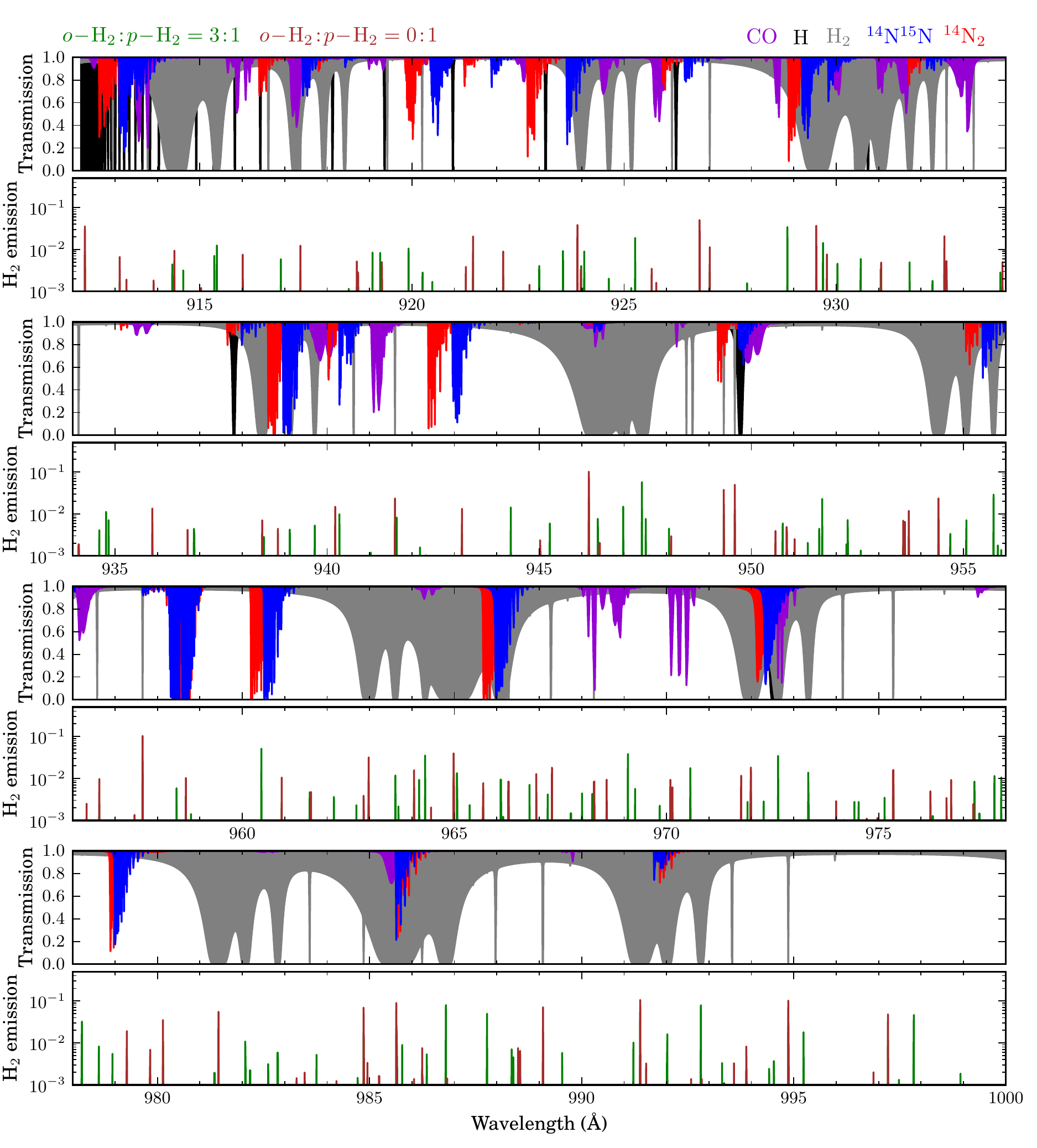}
  \caption{\emph{Upper subfigures:} Representative transmission functions calculated from theoretical photoabsorption cross sections of important molecular species, with column densities: $N(\mathrm{H})=10^{22}$, $N(\mathrm{H_2})=10^{20}$, and $N(\mathrm{{}^{14}N_2})=N(\mathrm{{}^{14}N^{15}N})=N(\mathrm{CO})=10^{15}\,\mathrm{cm}^{-2}$; and simulating an excitation temperature of 50\,K for all molecular absorption. \emph{Lower subfigures:} Probability per \AA{} of H$_2$ emission following a cosmic-ray induced ionisation event. Spectra are shown assuming two different values for the H$_2$ ortho:para ratio.}
  \label{fig:cross sections}
\end{figure*}

We have calculated \ce{^{14}N ^{15}N} photoabsorption and photodissociation cross sections as \citet{li2013} did for \ce{^{14}N2}.
A comparison is made in Fig.~\ref{fig:cross sections} between transmission functions derived from the two CSE-calculated photoabsorption cross sections.
The principal difference is a shift of \ce{^{14}N ^{15}N} absorption bands to longer wavelengths which mostly increases with vibrational excitation, that is, with shortening wavelength.
However, there are many further differences between the spectra of different N$_2$ isotopologues due to the strong coupling between excited states, such as depicted in Fig.~\ref{fig:b01 predissoc frac} with respect to the $b\,{}^1\Pi_u(v=1)$ predissociation fraction, and these should not be neglected \citep[e.g.,][]{vieitez_etal2008a,lewis_etal2008a,heays2011b}.

The wavelengths of the majority of the modelled rovibrational lines are accurate to within 0.01\,\AA{} when compared with laboratory measurements \citep{heays2011}. 
The magnitudes of the calculated cross sections are accurate to about 10\%, with their principal uncertainty arising from the absolute calibration of the laboratory absorption oscillator strengths used to constrain the model electronic transition moments \citep{stark_etal2005,stark_etal2008,heays_etal2009}.
The predissociation fractions pertaining to the longest wavelength bands are well known and modelled \citep{lewis_etal2005a} but are less certain at shorter wavelengths.
For these, a predissociation fraction of 1 is assumed.
This is a reliable assumption with respect to ${}^1\Pi_u$ states, many of which are known to exhibit significant predissociation broadening \citep{stark_etal2008,heays2011b}.
It is less certain that all ${}^1\Sigma^+_u$ levels responsible for absorption at shorter wavelengths are completely predissociative.
Further practical details regarding the calculation of N$_2$ cross sections and dissociation fractions at different temperatures may be found in \citet{li2013}.

Also shown in Fig.~\ref{fig:cross sections} are transmission functions representing further line-absorbing species relevant to the following treatment of the shielding of N$_2$ absorption in astrophysical environments.
For these, photoabsorption cross sections for H$_2$ are synthesised from data obtained from the Meudon PDR code website \citep{le_petit2006}, and those for CO are taken from the model of \citet{visser2009}.

\section{Photodissociation rates and shielding functions}
\label{sec:pd rates}
\label{sec:shielding functions}

The photodissociation rate, $k_0$, of N$_2$ exposed to UV radiation can be calculated according to
\begin{equation}
  \label{eq:kpd}
 k_0 = \int \sigma^\mathrm{pd} (\lambda) I(\lambda)d \lambda, 
\end{equation}
where $\sigma^\mathrm{pd}(\lambda)$ is the N$_2$ photodissociation cross section described in Sec.~\ref{sec:CSE model}, $I$ is the intensity of the radiation field, and $\lambda$ the wavelength.
\change{The unattenuated interstellar radiation field (ISRF) of \citet{draine1978}, in units of photons cm$^{-2}$\,s$^{-1}$\,\AA{}$^{-1}$\,sr$^{-1}$, is used in most of the following calculations and is given by
\begin{multline}
\label{eq:draine field differential}
 F(\lambda) = \frac{1}{4\pi} ( 3.2028 \times 10^{15} \lambda^{-3} - 5.1542 \times 10^{18} \lambda^{-4} \\ + 2.0546 \times 10^{21} \lambda^{-5} ) ,
\end{multline}
with $\lambda$ in \AA.
In the following we employ an angle-integrated intensity,
\begin{equation}
I(\lambda) = 4\pi \chi F(\lambda),
\label{eq:draine field}
\end{equation}
with an additional scaling factor relative to the Draine field, $\chi$.
}
The integral in Eq.~(\ref{eq:kpd}) must be computed between 912 and 1000\,\AA{}.
This range is defined by the long-wavelength onset of the N$_2$ absorption spectrum and the complete suppression of ultraviolet radiation for wavelengths shorter than the atomic H ionisation limit.

The photodissociation rates of \ce{^{14}N2} and $\mathrm{{}^{14}N{}^{15}N}$ calculated from Eq.~(\ref{eq:kpd}) are given in Tab.~\ref{tab:kpd0} assuming several different radiation fields.
In all cases the difference between \ce{^{14}N2} and $\mathrm{{}^{14}N{}^{15}N}$ rates is below 5\%, and the calculated dependences on the N$_2$ ground state excitation temperature and the form of the incident field (a Draine ISRF, or blackbody radiation) are similar to those calculated by \citet{li2013} for \ce{^{14}N2}.

For N$_2$ embedded in an externally irradiated cloud or disc, the intervening gaseous and granular material acts as a shield \change{and leads to a reduced photodissociation rate, $k<k_0$.}
The precise nature of shielding by H, H$_2$, \ce{^{14}N2}, and $\mathrm{{}^{14}N{}^{15}N}$ is influenced by the wavelength dependence of their absorption cross sections, but is more usefully applied as a wavelength-integrated ratio of shielded and unshielded dissociation rates.
That is, the shielding function given by
\begin{align}
  \notag
  \theta &= \frac{k}{k_0} \\
  \label{eq:def shielding function}
  &= \frac{
    \int I(\lambda)
    \exp\left[-\sum_\mathrm{X} N_\mathrm{X}\sigma^\mathrm{abs}_\mathrm{X}(\lambda)\right]
    \exp\left( -\gamma_\mathrm{dust} A_V \right)
    \sigma^\mathrm{pd}(\lambda)\,d\lambda}
  {\int I(\lambda)\sigma^\mathrm{pd}(\lambda)\,d\lambda}.
\end{align}
Here, $\sigma_\mathrm{X}^\mathrm{abs}(\lambda)$, is the absorption cross section of shielding species X and $N_\mathrm{X}$ its shielding column density.

\begin{table}
  \centering
  \caption{The unattenuated photodissociation rates ($k_0$, s$^{-1}$) of N$_2$ \change{exposed to a completely-surrounding Draine interstellar radiation field for various excitation temperatures ($T_\mathrm{ex}$)}, and for blackbody radiation with temperature $T_{\rm bb}$ and $T_\mathrm{ex}=30$\,K.}
  \label{tab:kpd0}
  \begin{tabular}{rcc}
    \hline\hline
    \\[-2ex]
    \multicolumn{3}{c}{Draine ISRF ($\chi=1$)}\\[1ex]
    $T_\mathrm{ex}$(\,K) & \ce{^{14}N2}                 & $\mathrm{{}^{14}N{}^{15}N}$\\
    \hline
    \\[-2ex]
    10                   & $1.64\times 10^{-10}$ & $1.68\times 10^{-10}$\\
    20                   & $1.65\times 10^{-10}$ & $1.70\times 10^{-10}$\\
    30                   & $1.66\times 10^{-10}$ & $1.71\times 10^{-10}$\\
    50                   & $1.67\times 10^{-10}$ & $1.72\times 10^{-10}$\\
    100                  & $1.70\times 10^{-10}$ & $1.74\times 10^{-10}$\\
    1000                 & $1.79\times 10^{-10}$ & $1.87\times 10^{-10}$\\
    \hline
    \\[2ex]
    \hline\hline
    \\[-2ex]
    \multicolumn{3}{c}{Blackbody radiation\tablefootmark{a} \ \
 ($T_\mathrm{ex}=30\,\mathrm{K}$)}\\[1ex]
    $T_{\rm bb}$(\,K)    & \ce{^{14}N2}                 & $\mathrm{{}^{14}N{}^{15}N}$\\
    \hline
    \\[-2ex]
    4000                 &$2.35\times 10^{-16}$ & $2.44\times 10^{-16}$\\
    6000                 &$1.04\times 10^{-13}$ & $1.07\times 10^{-13}$\\
    8000                 &$1.95\times 10^{-12}$ & $2.01\times 10^{-12}$\\
    10\,000              &$1.05\times 10^{-11}$ & $1.08\times 10^{-11}$\\
    20\,000              &$1.98\times 10^{-10}$ & $2.02\times 10^{-10}$\\
    \hline
  \end{tabular}
  \tablefoot{\tablefoottext{a}{The integrated \change{energy} intensities of all radiation fields over the interval 912--2050\,\AA{} have been normalised to a $\chi=1$ \citet{draine1978} field.}}
\end{table}

The term $\exp\left( -\gamma_\mathrm{dust} A_V \right)$ in Eq.~(\ref{eq:def shielding function}) describes shielding of the UV flux by dust grains, where the visual (5500\,\AA) extinction is assumed proportional to the column density of hydrogen nuclei, $N_\mathrm{H}$, according to $A_V = N_\mathrm{H}/1.6\times 10^{21}$ \citep{savage1977}.
In principle, this is a wavelength dependent quantity which depends on the distribution of dust grain sizes, their composition, and their geometries.
The use of a simple parameterisation of declining intensity with increasing visual extinction is warranted by the significant uncertainties in these parameters.
Radiative transfer calculations considering the destruction and scattering of photons by a realistic distribution of interstellar dust grains were made by \citet{roberge1981,roberge1991}.
The ultraviolet extinction relevant to various molecules was parameterised by \citet{van_dishoeck2006} as a decaying exponential like that in Eq.~(\ref{eq:def shielding function}) using the the dust optical properties of \citet{roberge1991}.
Their extinction calculated for the wavelength range appropriate to CO is adopted here for \ce{N2}, that is with $\gamma_\mathrm{dust} = 3.53$.
Further calculations were made by \citet{van_dishoeck2006} assuming larger dust grains ($>1\mu$m in size) as identified in protoplanetary discs \citep{li2003a,jonkheid2006}. For this case $\gamma_\mathrm{dust}=0.6$.

\begin{figure*}
  \centering
  \begin{tabular}{cc}
    \includegraphics{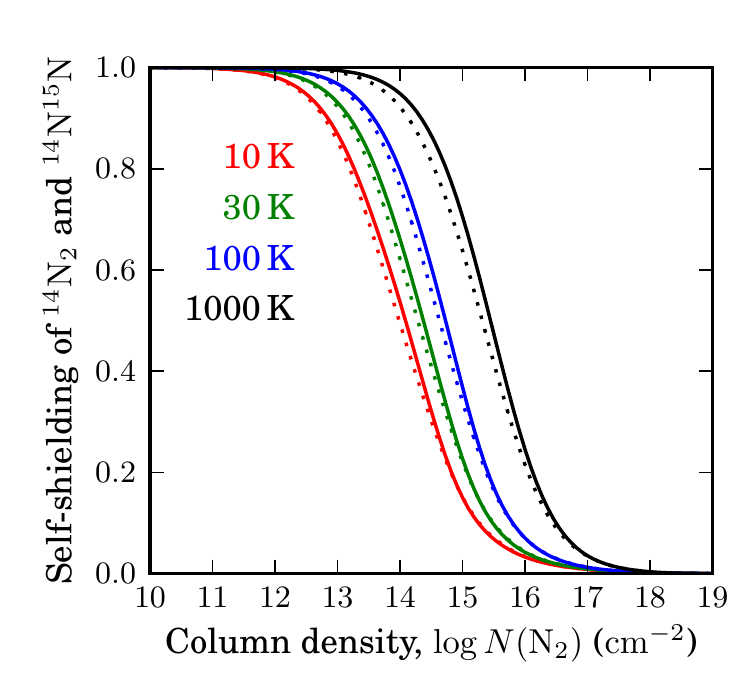} &
    \includegraphics{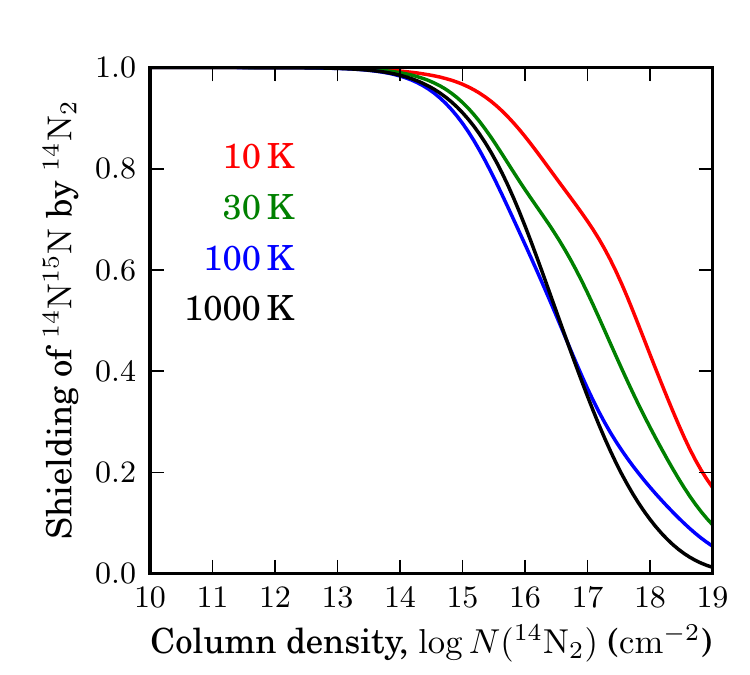} \\
    \multicolumn{2}{c}{\includegraphics{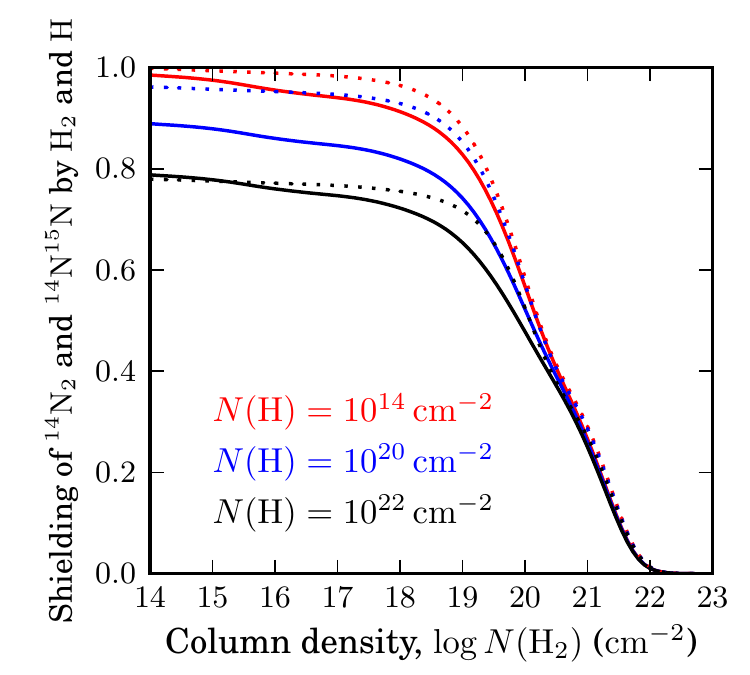}}
  \end{tabular}
  \caption{\emph{Top left:} Self-shielding functions for $\mathrm{{}^{14}N{}^{15}N}$ (\emph{solid lines}) and \ce{^{14}N2} (\emph{dotted lines}) at various temperatures defining their excitation state and thermal Doppler broadening. \emph{Top right:} Shielding of \ce{^{14}N ^{15}N} photodissociation by \ce{^{14}N2} at various excitation and Doppler broadening temperatures. \emph{Bottom:} Shielding of $\mathrm{{}^{14}N{}^{15}N}$ (\emph{solid lines}) and \ce{^{14}N2} (\emph{dotted lines}) photodissociation by H and H$_2$ as a function of their column densities.}
  \label{fig:shielding functions}
\end{figure*}
Detailed shielding functions for \ce{^{14}N2} were given in \citet{li2013} and the associated on-line material. 
Here, new calculations are made for $\mathrm{{}^{14}N{}^{15}N}$ using Eq.~(\ref{eq:def shielding function}), with some representative results plotted in Fig.~\ref{fig:shielding functions}, and full details available from the Leiden photodissociation database.\footnote{\tt www.strw.leidenuniv.nl/\textasciitilde{}ewine/photo}
The data in Fig.~\ref{fig:shielding functions} assume that the lines of shielding H and H$_2$ are Doppler broadened by $b=1$ and $3\,\mathrm{km\,s^{-1}}$, respectively.
The N$_2$ lines are assigned thermal Doppler widths according to their excitation temperatures but are also naturally broadened by the predissociation process, in some cases by a few \ce{km\,s^{-1}}.

Unsurprisingly, the similarity of the two isotopologues leads to very similar shielding functions.
There is some difference evident in Fig.~\ref{fig:shielding functions} with respect to shielding by atomic hydrogen. 
That is, the shielding of $\mathrm{{}^{14}N{}^{15}N}$ by a $N(\mathrm{H})=10^{20}\,\mathrm{cm}^{-2}$ column is somewhat more effective than for \ce{^{14}N2}.
This is the result of a nearer alignment of the $b\,{}^1\Pi_u(v=3)\leftarrow X\,{}^1\Sigma_g^+(v''=0)$ absorption band of \ce{^{14}N2} and the line-centre of the shielding $4p\leftarrow 1s$ transition of atomic hydrogen.
The relevant \ce{^{14}N2}, \ce{^{14}N ^{15}N}, and H features are plotted in Fig.~\ref{fig:cross sections} near 972\,\AA.
For a larger H column, like that in Fig.~\ref{fig:cross sections}, with heavily saturated and broadened H lines, the misalignment of H and \ce{^{14}N2} transitions near 972\,\AA{} loses its significance.

Also shown in Fig.~\ref{fig:shielding functions} is the shielding of \ce{^{14}N ^{15}N} photodissociation by \ce{^{14}N2}.
This rivals \ce{^{14}N ^{15}N} self-shielding because of the much greater abundance of \ce{^{14}N2}.
This is particularly true at higher temperatures where the wavelength offset between isotopologues is compensated for by their broader rotational distributions and Doppler widths.
A discussion of the relative importance of different species in mixed shielding media is given in Sec.~\ref{sec:photochemical models} with respect to an example protoplanetary disc.

\change{
Shielding functions calculated according to Eq.~(\ref{eq:def shielding function}) describe the reduction in photodissociation rate due to the attenuation of unidirectional radiation.
A related quantity, $\theta_\text{iso}(N)$, describes this reduction as a function of column density for an extended medium illuminated isotropically at its surface.
The resultant non-perpendicular rays penetrate less deeply and $\theta_\text{iso}(N)$ is smaller than the shielding function in the unidirectional case, $\theta(N)$ \citep{roellig2007}.
For a plane-parallel shielding medium with $N$ defined perpendicular to its surface, the two shielding functions are related geometrically by the formula
\begin{equation}
  \label{eq:semi-infinit isotropic shielding functions}
  \theta_\text{iso}(N) =  \int_0^{\frac{\pi}{2}} \theta\left(\frac{N}{\cos \phi}\right)\sin\phi\,d\phi.
\end{equation}
}

\section{Photodissociation due to cosmic-rays}
\label{sec:CR}

\subsection{Cosmic-ray induced emission and photodissociation model}

The ionisation of H$_2$ by cosmic ray collisions leads to the generation of free electrons.
These may collisionally excite (or ionise) further H$_2$ whose radiative relaxation generates an ultraviolet flux \citep{prasad1983,cecchi-pestellini1992b}.
\citet{sternberg1987}, \citet{gredel1987}, and \citet{gredel1989} estimated the photodissociation properties of several small molecules due to ultraviolet photons arising from this process.
Similar calculations are made here for the dissociation of N$_2$ including the detailed effects of line absorption as modelled by the CSE dissociation cross sections.

\change{
The H$_2$ emission flux between 912 and 1000\,\AA{} is dominated by line transitions. 
This emission occurs primarily through the Lyman and Werner bands connecting the ground state with the $B\,{}^1\Sigma_u^+$ and $C\,{}^1\Pi_u$ excited states \citep{sternberg1987}, and with a smaller contribution arising from other excited states \citep{gredel1989}.
Here, we use a model simulating the excitation of 9 electronic states of \ce{H2} by cosmic-ray generated electrons and the resulting emission, developed and expounded in detail by \citet{gredel1987} and \citet{gredel1989}.
This model includes the effects of cascading from sufficiently high-energy states into $B\,{}^1\Sigma_u^+$ and competing predissociative decay.
}

\change{
The final result of the H$_2$ emission model is a spectrum, $P(\lambda)$, describing the number of photons generated per unit spectral density per H$_2$ ionisation event.
A plot of $P(\lambda)$ is shown in Fig.~\ref{fig:cross sections} and demonstrates a large number of narrow emission lines which actually occupy very little of the wavelength interval shown.
The widths of these lines are Doppler limited and assigned full-width half-maxima of 0.005\,\AA{}, which is appropriate for a typical dark-cloud non-thermal Doppler broadening of $b=1\,\mathrm{km\,s^{-1}}$. 
This is comparable to the widths of N$_2$ absorption lines, also shown in Fig.~\ref{fig:cross sections}, so model spectra of both species are treated at similarly high resolution.
Spectra are shown in Fig.~\ref{fig:cross sections} assuming limiting high- and low-temperature ratios for the ortho:para H$_2$ populations, $o\!-\!\mathrm{H}_2\!:\!p\!-\!\mathrm{H}_2=3:1$ and $0:1$, respectively.
The difference is quite apparent and is significant in the following calculations of N$_2$ photodissociation.
Despite the 170\,K energy separation of $o\!-\!\mathrm{H}_2$ and $p\!-\!\mathrm{H}_2$ ground-state levels it is not evident that their low-temperature equilibrium populations are always attained in real molecular clouds \citep{tielens2013}. 
Thus, we choose to contrast high and low temperature limits in the present study.
}

\change{The rate of ultraviolet photons generated by cosmic ray collisions per unit spectral density per hydrogen nucleus is given by
\begin{equation}
  \label{eq:CR UV intensity}
  R(\lambda) = \zeta x_\mathrm{H_2} P(\lambda),
\end{equation}
where $x_\mathrm{H_2}= n(\mathrm{H}_2) / \left[ n(\mathrm{H})+2n(\mathrm{H}_2) \right]$ is the relative abundance of molecular hydrogen with respect to H nuclei and  $\zeta$ is the cosmic-ray-induced ionisation rate per H$_2$ molecule.}
A value of $\zeta=10^{-16}\ \text{s}^{-1}$ per \ce{H2} is adopted for the following calculations and falls within the broad range of rates estimated from observations of \change{ions in diffuse and translucent clouds} \citep{dalgarno2006,padovani2009,indriolo2012,rimmer2012}.
This collection of observations suggest a rate as high as $10^{-15}\,\mathrm{s}^{-1}$ at the edge of an interstellar cloud, which decreases with extinction to $\mathord{\sim}10^{-16}\,\mathrm{s}^{-1}$ for $A_V=10$\,mag as the lowest-energy cosmic rays are attenuated.
Our adopted value is higher than observed in the interiors of the densest clouds with $\zeta < 10^{-17}\,\mathrm{s}^{-1}$ \citep{hezareh2008}.
Theoretical consideration of a range of source and attenuation mechanisms influencing the cosmic ray flux  \change{\citep{padovani2009,cleeves2013}} suggest a smaller rate $\zeta$ than $10^{-16}\,\mathrm{s}^{-1}$ may be expected in circumstellar environments, because of the possible influence of any stellar wind and magnetic field.
The photodissociation results presented here may be adapted to ionisation rates other than $\zeta=10^{-16}\,\mathrm{s}^{-1}$ by a simple scaling.

The cosmic-ray generated photons may be absorbed by dust grains and by abundant line-absorbing species: H$_2$, H, N$_2$, and CO.
The idea of an absorbing column density is inappropriate because of the local nature of the photon source.
Instead, the fraction of ultraviolet radiation absorbed by a particular species, X, and leading to its dissociation is determined from its relative abundance $x_\text{X}$, according to 
\begin{equation}
  \label{eq:absorption fraction}
  p_\text{X}(\lambda) = \frac{x_\text{X}\sigma^\mathrm{pd}_\text{X}(\lambda)}{x_\mathrm{dust}\sigma_\mathrm{dust} + \sum_i x_i\sigma^\mathrm{abs}_i(\lambda)}.
\end{equation}
Here, $x_i$ and $\sigma_i^\mathrm{abs}(\lambda)$ are the abundances and photoabsorption cross sections of various species, $x_\mathrm{dust}\sigma_\mathrm{dust}$ represents dust absorption, and $\sigma^\mathrm{pd}_\text{X}(\lambda)$ is a photodissociation cross section.

An important term in the summation in Eq.~(\ref{eq:absorption fraction}) arises from H$_2$.
For the cases of X being \ce{^{14}N2} or CO, self-shielding is important and must also be included.
Self-shielding and shielding by \ce{^{14}N2} are both important for the case of \ce{^{14}N ^{15}N} photodissociation.
 Atomic hydrogen is also a source of UV shielding at the edge of a cloud or disc but is assumed here to have too low abundance (about $10^{-4}$ relative to H$_2$) to be important in fully shielded regions. 
Consequently, the common assumption that all radiation shortwards of 912\,\AA{}{} is completely consumed by H ionisation does not apply here. 
However, the N$_2$ photodissociation cross section at shorter wavelengths and approaching the photoionisation limit at 800\,\AA{} contributes less than 10\% to its total photodissociation rate, and is neglected here.
 
Shielding of a penetrating interstellar radiation field by dust grains was discussed in Sec.~\ref{sec:shielding functions}.
The extinction of cosmic-ray induced radiation is treated differently.
In this case, all photons are assumed absorbed before traversing a significant fraction of the dark interior of the interstellar cloud or disk where they are generated.
Then, the detailed angular scattering of photons by dust grains is inconsequential and only the absorption cross section need be considered.
The dust absorption cross section is related to its extinction cross section by 
\begin{equation}
  \label{eq:dust abs ext}
  \sigma^\text{abs}_\text{dust} = \sigma^\text{ext}_\text{dust}(1-\omega),
\end{equation}
where $\omega$ is the grain albedo.
We adopt a commonly used value derived from observations \citep{savage1977,bohlin1978} for the dust extinction cross section per hydrogen nucleus of $2\times 10^{-21}\,\mathrm{cm}^2$ and assume an albedo of $\omega=0.5$.
Then, $x_\mathrm{dust}\sigma^\text{abs}_\mathrm{dust} = 10^{-21}\,\mathrm{cm^2}$ in Eq.~(\ref{eq:absorption fraction}).

\change{The assumption of locally-absorbed UV photons is only appropriate for a sufficiently large and homogeneous medium.
The mean-free-path of cosmic-ray generated photons in an interstellar cloud of density $n_\text{H}=10^3$\,cm$^{-3}$ is 30\,000\,AU, assuming the UV-to-visual extinction factor of 3.53 discussed in Sec.~\ref{sec:shielding functions}.
For the larger densities found near the midplane of a protoplanetary disc \citep{jonkheid2006} this distance is accordingly reduced, e.g., $n_\text{H}=10^8$\,cm$^{-3}$  and $\sim1$\,AU.
In principle, the various cosmic-ray induced photodissociation rates calculated below will be overestimated for objects which are not significantly larger than this mean-free-path, due to the escape of some UV photons.
However, the significance of cosmic-ray induced photon fluxes is generally limited to regions fully-shielded from external UV radiation, and such regions also necessarily shield the escape of internally-generated photons.
}

The dust grains in a protoplanetary disc are likely larger than those in interstellar space, which have radii $<1\,\mu\mathrm{m}$ \citep{mathis1977}.
\citet{li2003a} deduced properties of the dust grains in one protoplanetary system, HD\,4796A, by reference to observed infrared and sub-millimetre wavelength emission.
Their modelling required a lower limit of 1\,$\mu$m on the grain size distribution.
The N$_2$-dissociating cosmic-ray induced UV flux considered here has wavelengths shorter than 1000\,\AA{} and will not be as strongly absorbed by grains larger than this.
However, a population of polycyclic aromatic hydrocarbons (PAHs) has been inferred in a number of discs \citep[e.g.,][]{li2003b,geers2006,geers2007,maaskant2013} and will absorb extra UV radiation \citep{siebenmorgen}. 
\citet{chaparro_molano2012a} assumed a dust distribution with a minimum grain size of 0.1\,$\mu$m in their study of cosmic-ray induced photodissociation in a model disc. Following from this assumption they deduced a value of $x_\mathrm{dust}\sigma^\text{abs}_\mathrm{dust} = 1.47\times 10^{-22}\,\mathrm{cm^2}$, significantly below the commonly-adopted interstellar value.
Here, we explore an even lower value, $x_\mathrm{dust}\sigma^\text{abs}_\mathrm{dust} = 10^{-23}\,\mathrm{cm^2}$, in order to contrast a more extreme case of grain growth in a protoplanetary disc with interplanetary dust.

Finally, the photodissociation rate of species X (per molecule X) due to cosmic ray induced photons may be calculated according to
\begin{equation}
  \label{eq:CR abs pd rate}
  k_\text{X} = \frac{1}{x_\mathrm{X}}\int R(\lambda)p_\mathrm{X}(\lambda)\,d\lambda.
\end{equation}
In some previous descriptions of cosmic-ray induced photodissociation \citep{gredel1989,mcelroy2013} an efficiency is defined by factoring the H$_2$ ionisation rate and dust grain albedo from Eq.~(\ref{eq:CR abs pd rate}).
In this case, the importance of molecular line absorption in Eq.~(\ref{eq:absorption fraction}) prevents the factoring of dust properties.

\subsection{Cosmic-ray induced photodissociation of \ce{N2}}
\label{sec:CR pd N2}

\begin{table}
  \centering
  \caption{Photodissociation rate per molecule due to cosmic-ray induced radiation.}
  \begin{tabular}{lrrr}
    \hline\hline\\[-1ex]
    Dissociation rate ($\times10^{-16}\,\mathrm{s^{-1}}$):                              & \multicolumn{1}{c}{\ce{^{14}N2}} & \multicolumn{1}{c}{\ce{^{14}N ^{15}N}} & \multicolumn{1}{c}{CO} \\
    \hline
    \\
    \multicolumn{4}{c}{Base conditions for an interstellar cloud} \\
    \\
    $\zeta=10^{-16}\,\mathrm{s^{-1}}$                                                   &                                  &                                        & \\
    $x_\mathrm{{}^{14}N_2}=10^{-5}$                                                     &                                  &                                        & \\
    $x_\mathrm{{}^{14}N^{15}N}=x_\mathrm{{}^{14}N_2}/225$                               &                                  &                                        & \\
    $x_\mathrm{CO}=7\times 10^{-5}$                                                     &                                  &                                        & \\
    $T_\mathrm{ex}=20\,\mathrm{K}$                                                      & 14\phantom{.0}                   & 21\phantom{.0}                         & 8.1 \\
    $b=1\,\mathrm{km\,s^{-1}}$                                                          &                                  &                                        & \\
    $o\!-\!\mathrm{H}_2\!:\!p\!-\!\mathrm{H}_2=3:1$                                     &                                  &                                        & \\
    $x_\mathrm{dust}\sigma^\text{abs}_\mathrm{dust} = 10^{-21}\,\mathrm{cm}^2$          &                                  &                                        & \\
    \\
    \\
    \multicolumn{4}{c}{Variations} \\
    \\
    $T_\mathrm{ex}=10\,\mathrm{K}$                                                      & 13\phantom{.0}                   & 44\phantom{.0}                         & 12\phantom{.0}\\
    $T_\mathrm{ex}=30\,\mathrm{K}$                                                      & 18\phantom{.0}                   & 28\phantom{.0}                         & 9.2\\
    $T_\mathrm{ex}=100\,\mathrm{K}$                                                     & 8.7\phantom{.0}                  & 30\phantom{.0}                         & 7.5\\
    $T_\mathrm{ex}=300\,\mathrm{K}$                                                     & 6.8                              & 20\phantom{.0}                         & 10\phantom{.0}\\
    \\
    $b=3\,\mathrm{km\,s^{-1}}$                                                          & 17\phantom{.0}                   & 11\phantom{.0}                         & 9.1\\
    $b=5\,\mathrm{km\,s^{-1}}$                                                          & 19\phantom{.0}                   & 12\phantom{.0}                         & 10\phantom{.0}\\
    \\
    $x_\mathrm{{}^{14}N_2}=10^{-8}$, $x_\mathrm{CO}=7\times10^{-8}$                     & 47\phantom{.0}                   & 36\phantom{.0}                         & 54\phantom{.0}\\
    $x_\mathrm{{}^{14}N_2}=10^{-7}$, $x_\mathrm{CO}=7\times10^{-7}$                     & 45\phantom{.0}                   & 36\phantom{.0}                         & 40\phantom{.0}\\
    $x_\mathrm{{}^{14}N_2}=10^{-6}$, $x_\mathrm{CO}=7\times10^{-6}$                     & 33\phantom{.0}                   & 33\phantom{.0}                         & 21\phantom{.0}\\
    $x_\mathrm{{}^{14}N_2}=10^{-4}$, $x_\mathrm{CO}=7\times10^{-4}$                     & 2.8                              & 8.4                                    & 2.1\\
    \\
    $T_\mathrm{ex}=10\,\mathrm{K}$, $o\!-\!\mathrm{H}_2\!:\!p\!-\!\mathrm{H}_2=0\!:\!1$ & 8.0                              & 6.6                                    & 8.0\\
    $T_\mathrm{ex}=20\,\mathrm{K}$, $o\!-\!\mathrm{H}_2\!:\!p\!-\!\mathrm{H}_2=0\!:\!1$ & 8.9                              & 22\phantom{.0}                         & 9.5\\
    $T_\mathrm{ex}=30\,\mathrm{K}$, $o\!-\!\mathrm{H}_2\!:\!p\!-\!\mathrm{H}_2=0\!:\!1$ & 9.2                              & 42\phantom{.0}                         & 10\phantom{.0}\\
    \\
    $x_\mathrm{{}^{14}N^{15}N}=x_\mathrm{{}^{14}N_2}/100$                               & \multicolumn{1}{c}{--}           & 17\phantom{.0}                         & \multicolumn{1}{c}{--}\\
    $x_\mathrm{{}^{14}N^{15}N}=x_\mathrm{{}^{14}N_2}/50$                                & \multicolumn{1}{c}{--}           & 14\phantom{.0}                         & \multicolumn{1}{c}{--}\\
    \\
    \\
    \multicolumn{4}{c}{Modification for a protoplanetary disc} \\
    \multicolumn{4}{c}{$x_\mathrm{dust}\sigma^\text{abs}_\mathrm{dust} = 10^{-23}\,\mathrm{cm}^2$} \\
    \\
    $T_\mathrm{ex}=10\,\mathrm{K}$, $o\!-\!\mathrm{H}_2\!:\!p\!-\!\mathrm{H}_2=3\!:\!1$ & 18\phantom{.0}                   & 69\phantom{.0}                         & 19\phantom{.0}\\
    $T_\mathrm{ex}=20\,\mathrm{K}$, $o\!-\!\mathrm{H}_2\!:\!p\!-\!\mathrm{H}_2=3\!:\!1$ & 21\phantom{.0}                   & 52\phantom{.0}                         & 15\phantom{.0}\\
    $T_\mathrm{ex}=30\,\mathrm{K}$, $o\!-\!\mathrm{H}_2\!:\!p\!-\!\mathrm{H}_2=3\!:\!1$ & 23\phantom{.0}                   & 75\phantom{.0}                         & 16\phantom{.0}\\
    \\
    $T_\mathrm{ex}=10\,\mathrm{K}$, $o\!-\!\mathrm{H}_2\!:\!p\!-\!\mathrm{H}_2=0\!:\!1$ & 11\phantom{.0}                   & 18\phantom{.0}                         &17\phantom{.0}\\
    $T_\mathrm{ex}=20\,\mathrm{K}$, $o\!-\!\mathrm{H}_2\!:\!p\!-\!\mathrm{H}_2=0\!:\!1$ & 12\phantom{.0}                   & 62\phantom{.0}                         &18\phantom{.0}\\
    $T_\mathrm{ex}=30\,\mathrm{K}$, $o\!-\!\mathrm{H}_2\!:\!p\!-\!\mathrm{H}_2=0\!:\!1$ & 14\phantom{.0}                   & 118\phantom{.0}                        &17\phantom{.0}\\
    \\
    \hline
  \end{tabular}
  \label{tab:cosmic ray photodissociation rates}
\end{table}
The rate of dissociation of \ce{^{14}N2} and \ce{^{14}N ^{15}N} due to cosmic-ray induced ultraviolet radiation has been calculated and listed in Tab.~\ref{tab:cosmic ray photodissociation rates}.
This rate is dependent on the local physical conditions, and a set of base conditions is chosen which plausibly simulate the interior of a dark interstellar cloud:\
\begin{itemize}
\item[$\bullet$] The assumed abundances of \ce{^{14}N2} and CO relative to $n_\mathrm{H}=n(\ce{H})+\change{2}n(\ce{H2})$ are $10^{-5}$ and $7\times10^{-5}$, respectively, and reflect typical dense cloud abundances \citep{aikawa2008}.
\item[$\bullet$] A $\mathrm{{}^{14}N_2}\!:\!\mathrm{{}^{14}N{}^{15}N}$ ratio of 225 is chosen, appropriate for the solar elemental ratio of 450 \citep{marty2010}. 
\item[$\bullet$] All molecular species are assumed to have a thermal excitation of $T_\mathrm{ex}=20\,\mathrm{K}$ and be turbulently broadened with Doppler velocity $b=1\,\mathrm{km\,s^{-1}}$.
\item[$\bullet$] Dust grains are assumed to have a scattering cross section per H nucleus of $x_\mathrm{dust}\sigma^\text{abs}_\mathrm{dust} = 10^{-21}\,\mathrm{cm^2}$.
\item[$\bullet$] The high temperature limit, 3:1, is assumed for the ortho:para ratio of emitting H$_2$.
\end{itemize}
Further calculations listed in Tab.~\ref{tab:cosmic ray photodissociation rates} probe variations on this set of parameters.

For the base set of parameters considered in Tab.~\ref{tab:cosmic ray photodissociation rates}, the \ce{^{14}N2} photodissociation rate is $\mathrm{14\times10^{-16}\,s^{-1}}$. 
This is 5 orders of magnitude below the rate calculated for a standard interstellar radiation field in Sec.~\ref{sec:pd rates}. 
The cosmic-ray induced photodissociation rates calculated by \citet{gredel1989} for a large number of other molecular species are generally larger than the \ce{N2} value determined here, in many cases by multiple orders of magnitude.
For example, assuming $\zeta=10^{-16}\,\mathrm{s}^{-1}$ and a grain albedo of $\omega=0.5$, their photodissociation rates for \ce{H2O} and \ce{CN} are $2.0\times 10^{-14}$ and $1.8\times 10^{-13}\,\mathrm{s}^{-1}$, respectively.

The self-shielding of ${}^{14}$N$_2$ cannot be neglected because of the resonant nature of its spectrum.
For this reason, decreasing the \ce{^{14}N2} density rapidly increases the dissociation rate, until the abundance relative to H nuclei, $x_\mathrm{{}^{14}N_2}$, falls below about $10^{-7}$, as is evident from the values in Tab.~\ref{tab:cosmic ray photodissociation rates}.
Line absorption of H$_2$ also effectively shields N$_2$ dissociation, reducing its rate by a factor of 2.
Dust opacity reduces the ${}^{14}\mathrm{N}_2$ dissociation rate by a further 35\%, and shielding by CO less than 5\%.

\begin{figure*}
  \centering
  \begin{tabular}{c@{}c}
    \includegraphics{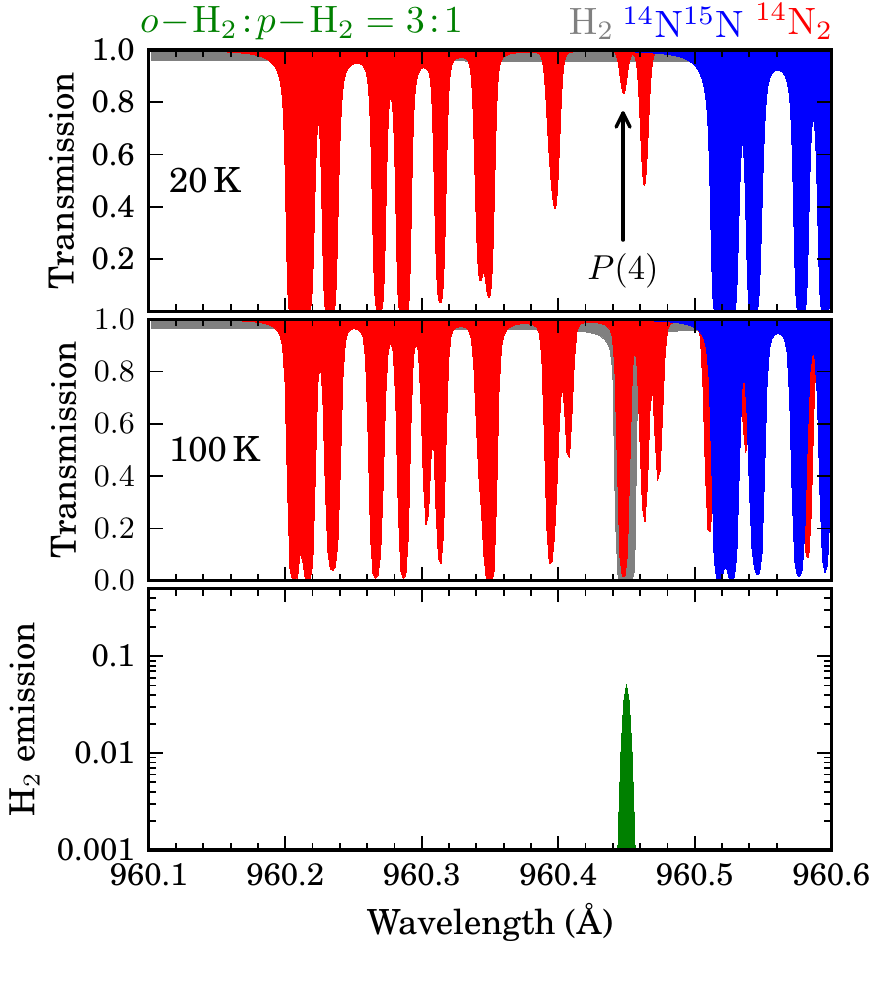} &
    \includegraphics{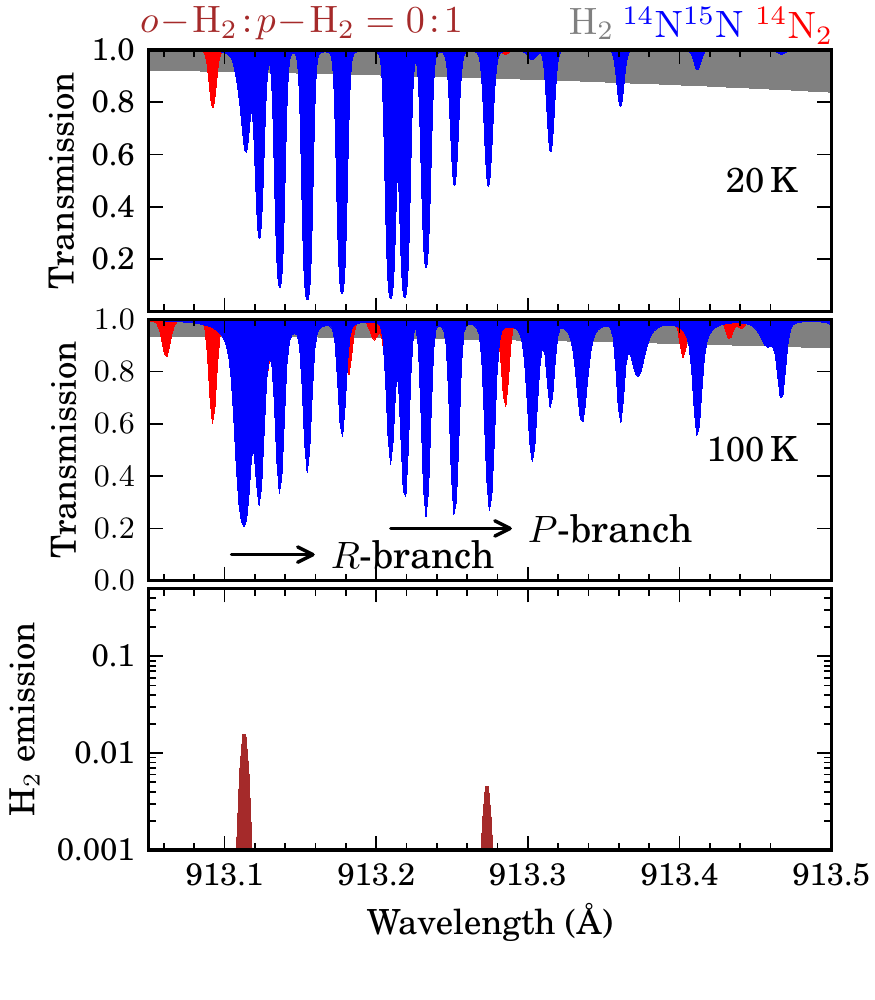} 
  \end{tabular}
  \caption{Modelled molecular absorption lines for two different excitation temperatures, 20 and 100\,K, and cosmic-ray induced \ce{H2} emission. Otherwise calculated as for Fig.~\protect\ref{fig:cross sections}. \emph{Left:} Overlap of an \ce{o - H2} emission line with the $c_3\,{}^1\Pi_u(v'=0)\leftarrow X\,{}^1\Sigma_g^+(v''=0)$ $P(4)$ absorption line of \ce{^{14}N2}.  \emph{Right:} Overlap of two \ce{p - H2} emission lines with absorption by the $o_3\,{}^1\Pi_u(v=2)\leftarrow X{}^1\Sigma_g^+(v=0)$ band of \ce{^{14}N ^{15}N}, with $R$ and $P$ branches indicated.}
  \label{fig:CR overlaps 960A and 913A}
\end{figure*}
\begin{figure}
  \centering
  \includegraphics{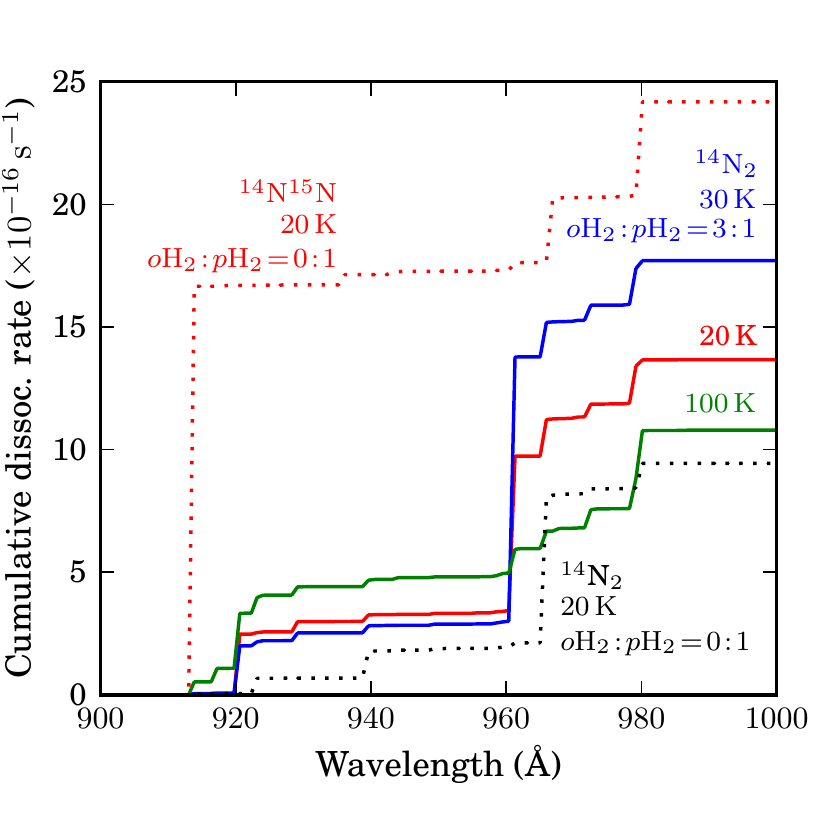}
  \caption{\change{The dissociation rate of N$_2$ due to cosmic-ray induced H$_2$ emission under interstellar cloud conditions, integrated cumulatively with increasing wavelength. Rates for \ce{^{14}N2} are plotted with $o\!-\!\mathrm{H}_2\!:\!p\!-\!\mathrm{H}_2=3\!:\!1$ for several temperatures \emph{(solid curves)}, for $o\!-\!\mathrm{H}_2\!:\!p\!-\!\mathrm{H}_2=3\!:\!1$ and $T_\text{ex}=20$\,K \emph{(black dotted curve)}, and a single case is included for \ce{^{14}N ^{15}N} \emph{(red dotted curve)}.}}
  \label{fig:14N2 cumulative CR dissoc rate}
\end{figure}

The effective dissociation of \ce{^{14}N2} arises from relatively few overlaps between its absorption lines and H$_2$ emission lines.
These are indicated by steps in an accumulation of the integrated dissociation rate with wavelength, as shown in Fig.~\ref{fig:14N2 cumulative CR dissoc rate}.
This accumulation is dominated by a line overlap occurring at 960.45\,\AA{}, plotted in detail in  Fig.~\ref{fig:CR overlaps 960A and 913A}.
This corresponds to the $P(4)$ line of the $c_3\,{}^1\Pi_u(v'=0)\leftarrow X\,{}^1\Sigma_g^+(v''=0)$ absorption band of \ce{^{14}N2} and the $P(3)$ line of the $B\,{}^1\Sigma_u^+(v'=13)\rightarrow X\,{}^1\Sigma_g^+(v''=0)$ H$_2$ Lyman emission band.
The significance of this overlap is dependent on the assumed excitation temperatures of absorbing \ce{^{14}N2} and shielding H$_2$.
When $T_\mathrm{ex}=20$\,K, also indicated in Figs.~\ref{fig:CR overlaps 960A and 913A} and \ref{fig:14N2 cumulative CR dissoc rate}, the fraction of N$_2$ ground state molecules in the $J=4$ rotational level is low and the significance of the 960.45\,\AA{} overlap is small.
For $T_\mathrm{ex}=100$\,K, the $J=3$ level population of ground state H$_2$ becomes significant and $P(3)$ absorption effectively shields \ce{^{14}N2} photodissociation.
Thus, the largest \ce{^{14}N2} photodissociation rate occurs at intermediate temperature.

Only small increases in the \ce{^{14}N2} photodissociation rate occur for increased turbulent broadening, $b$.
Greater sensitivity is found with respect to the assumed $o\!-\!\mathrm{H}_2\!:\!p\!-\!\mathrm{H}_2$ ratio.
This is limited to 0:1 and 3:1 at high and low temperatures, respectively, given sufficiently high densities for thermal equilibrium to be maintained.
The \ce{^{14}N2} photodissociation rate is approximately halved when the low temperature limit is adopted, primarily because the H$_2$ emission at 960.45\,\AA{}{} arises from an excited $o\!-\!\mathrm{H}_2$ level.

Table \ref{tab:cosmic ray photodissociation rates} also includes photodissociation rates assuming reduced absorption by dust grains, in order to simulate the interior of a protoplanetary disc.
The adopted value of $x_\mathrm{dust}\sigma^\text{abs}_\mathrm{dust}$ effectively removes dust from the competitive absorption of photons.
The N$_2$ photodissociation rate is then somewhat higher than for an interstellar cloud, but follows the same trends with respect to temperature and the $o\!-\!\mathrm{H}_2\!:\!p\!-\!\mathrm{H}_2$ ratio.
 
Calculations of the cosmic-ray induced photodissociation of $\mathrm{{}^{14}N{}^{15}N}$ are also presented in Tab.~\ref{tab:cosmic ray photodissociation rates}.
The resultant rate assuming the base conditions designed to represent a typical dark interstellar cloud is $21\times 10^{-16}\,\mathrm{s}^{-1}$.
This is slightly larger than for \ce{^{14}N2} and is the result of a different collection of overlapping N$_2$ absorption and H$_2$ emission lines.
The \ce{^{14}N ^{15}N} dissociation rate is found to be much larger than for \ce{^{14}N2} in the cold-temperature limit $o\!-\!\mathrm{H}_2\!:\!p\!-\!\mathrm{H}_2=0\!:\!1$ and for $T\simeq 30\,\mathrm{K}$. 
This is due to a strong overlap of H$_2$ emission lines near 913.1 and 913.3\,\AA{} with absorption lines corresponding to low rotational levels of the $o_3\,{}^1\Pi_u(v=2)\leftarrow X{}^1\Sigma_g^+(v=0)$ transition of \ce{^{14}N ^{15}N}, plotted in Fig.~\ref{fig:CR overlaps 960A and 913A}.
The influence of these overlaps dominate the accumulated dissociation rate shown in Fig.~\ref{fig:14N2 cumulative CR dissoc rate}.
The isotopologue difference between dissociation rates is increased significantly when the reduced dust shielding in a protoplanetary disc is accounted for and approaches a factor of 8.
Thus, from Tab.~\ref{tab:cosmic ray photodissociation rates}, the most likely candidate for isotopic fractionation and an enhancement of atomic ${}^{15}$N due to cosmic ray induced photodissociation occurs in disc material around 30\,K and with a small $o\!-\!\mathrm{H}_2\!/\!p\!-\!\mathrm{H}_2$ ratio.
The difference between \ce{^{14}N2} and \ce{^{14}N ^{15}N} dissociation rates diminishes for temperatures below 30\,K, and is negligible at 10\,K when assuming a low-temperature is $o\!-\!\ce{H2}\!/\!p\!-\!\ce{H2}$.

\subsection{Cosmic-ray induced photodissociation of CO}
\label{sec:CR pd CO}

We also combined the high-resolution \ce{H2} cosmic-ray induced emission model with the CO photodissociation cross sections of \citet{visser2009}.
There is an example CO transmission spectrum plotted in Fig.~\ref{fig:cross sections} which exhibits similar band structure as for \ce{N2}.
The reduced symmetry of heteronuclear CO allows for greater perturbation of its excited states and leads to more rapid predissociation on average than for \ce{N2}.
The CO rotational absorption lines are then somewhat broader than for \ce{N2}. 

The CO photodissociation rates in a cosmic-ray induced UV field assuming various physical parameters are listed in Tab.~\ref{tab:cosmic ray photodissociation rates} and are of a similar magnitude to the \ce{N2} rates. 
The on-average broader absorption lines of CO overlaps with \ce{H2} emission lines more frequently than for \ce{N2}, and the resulting photodissociation is then less sensitive to variations in the assumed excitation temperature and $o\!-\!\ce{H2}\!:\!p\!-\!\ce{H2}$ ratio.

The calculated CO photodissociation rate in Tab.~\ref{tab:cosmic ray photodissociation rates} for the base conditions representing an interstellar cloud is $8.1\times 10^{-16}$\,s$^{-1}$. 
This is somewhat less than the value calculated by \citet{gredel1987}, $13\times 10^{-16}$\,s$^{-1}$ (assuming $\omega=0.5$ and $\zeta=10^{-16}$\,s$^{-1}$), who used a less sophisticated model of CO photodissociation.

\section{Chemical models including photodissociation}
\label{sec:photochemical models}

\subsection{Interstellar cloud model}
\label{sec:interstellar cloud}

A chemical model simulating an interstellar cloud was studied in order to explore the details of $\mathrm{{}^{14}N{}^{15}N}$ photodissociation due to an interstellar radiation field impinging on its surface and a flux of cosmic rays within its interior.
The model is similar to that of \citet{li2013}.
Here, we used a modified version of the UMIST 2012 chemical network \citep{mcelroy2013}.
The network was stripped down to species containing only H, He, C, N and O; and a maximum of two C, N or O atoms.
Supra-thermal chemistry was included to enhance the formation of CH$^+$ (and thus also CO) at low $A_V$, following \citet{visser2009}. 
This boosts the rate of ion-neutral reactions after setting the Alfv{\'e}n speed to 3.3\,km\,s$^ {-1}$ for column densities less than $4\times 10^{20}$\,cm$^{-2}$.

Freeze-out and thermal evaporation for all neutral species were added to the gas-phase only UMIST network as well as the grain-surface hydrogenation reactions used by \citet{visser2011}.
The latter include H$_2$ formation and the conversion of C to \ce{CH4}, N to \ce{NH3}, O to \ce{H2O}, and S to \ce{H2S}.
\change{
Photodesorption due to the UV field generated from cosmic-ray ionisation and the attenuated ISRF were included, but direct cosmic-ray desorption was not.
}

\change{The wavelength dependence of the Draine ISRF was adopted for a unidirectional radiation field perpendicularly incident on the model cloud's edge.
This was scaled by a factor of \change{$\chi=1$} so that one-sided irradiation of the cloud leads to the same unshielded dissociation rates at its edge as those given in Tab.~\ref{tab:kpd0}.
The cloud is assumed sufficiently thick that no radiation penetrates from the far side.
Self-shielding of CO is computed using the shielding functions of \citet{visser2009}, for \ce{^{14}N2} we use those calculated by \citet{li2013} at 30\,K, and the presently calculated shielding functions are used for \ce{ ^{14}N ^{15}N}.
The scattering of UV radiation out of the incident beam is not included in our model.
}

\change{
The wavelength dependence of the Draine ISRF was designed to simulate an isotropic field generated from a remote stellar population \citep{draine1978}, and some previous interstellar cloud models also assume isotropic radiation \citep{roellig2007}.
The transition between atomic and molecular H$_2$ was shown by \citet{roellig2007} to differ between irradiation geometries, occurring at lower $A_V$ for the isotropic case relative to unidirectional radiation.
The transition for N$_2$ occurring in our unidirectional model would be similarly shifted to shallower depth if the incident radiation were isotropic.
}

The elemental abundances relative to H are those of \citet{aikawa2008}:  $0.0975$ for He, $7.86\times10^{-5}$ for C, $2.47\times10^{-5}$ for N, and $1.80\times10^{-4}$ for O.
The elemental ${}^{14}$N:${}^{15}$N ratio was fixed to the protosolar value, 450:1.
The rate of H$_2$ ionisation due to cosmic rays was set to $5\times 10^{-17}$\,s$^{-1}$.
The abundances of N, N$_2$ and CO reach steady state after $\mathord{\sim}$1\,Myr, regardless of whether the gas starts in atomic or molecular form.

The cloud was assumed to have a constant hydrogen-nuclei density of $n_\ce{H} = n(\mathrm{H})+2n(\mathrm{H}_2)=10^3\,\mathrm{cm}^{-3}$ and a temperature of 30\,K. 
The turbulent broadening, $b$, of CO, H$_2$, and H were set to 0.3, 3, and 5\,km\,s$^{-1}$; respectively; and N$_2$ lines were assigned linewidths equivalent to their thermal Doppler width at 30\,K, \change{in an identical fashion to the similar model of \cite{li2013}}.

The dust population was assigned a coefficient $\gamma_\mathrm{dust}=3.53$ in Eq.~(\ref{eq:def shielding function}) to simulate grains with an interstellar size distribution, with radius $\mathord{\sim}0.1\,\mu\textrm{m}$ size.
A value for the cosmic-ray induced photodissociation rate of N$_2$ was adopted corresponding to the low temperature $o\!-\!\mathrm{H}_2\!:\!p\!-\!\mathrm{H}_2$ ratio, 0:1. 
 
The nitrogen reaction network was augmented so that separate accounting of pure ${}^{14}$N and single-${}^{15}$N containing species was possible.
For this, all N-containing species and reactions were cloned and a ${}^{14}$N to ${}^{15}$N substitution made.
\change{
Two kinds of isotope-differentiating reactions are included in our model chemical network: photodissociation and low-temperature ion-molecule exchange reactions.
The selective effects of photodissociation are encoded in the shielding functions and cosmic-ray induced photodissociation rates presented in Secs.~\ref{sec:shielding functions} and \ref{sec:CR pd N2}.
Exchange reactions were added to our network using the rates listed in Tab.~2 of \citet{terzieva2000}.
These reactions exothermically favour increased abundances of \ce{^{15}N} in more complex species because of the lower zero-point energy of molecules with heavier nuclei, with typical excess energies of 10 to 30\,K.
}

\begin{figure*}
  \centering
  \includegraphics{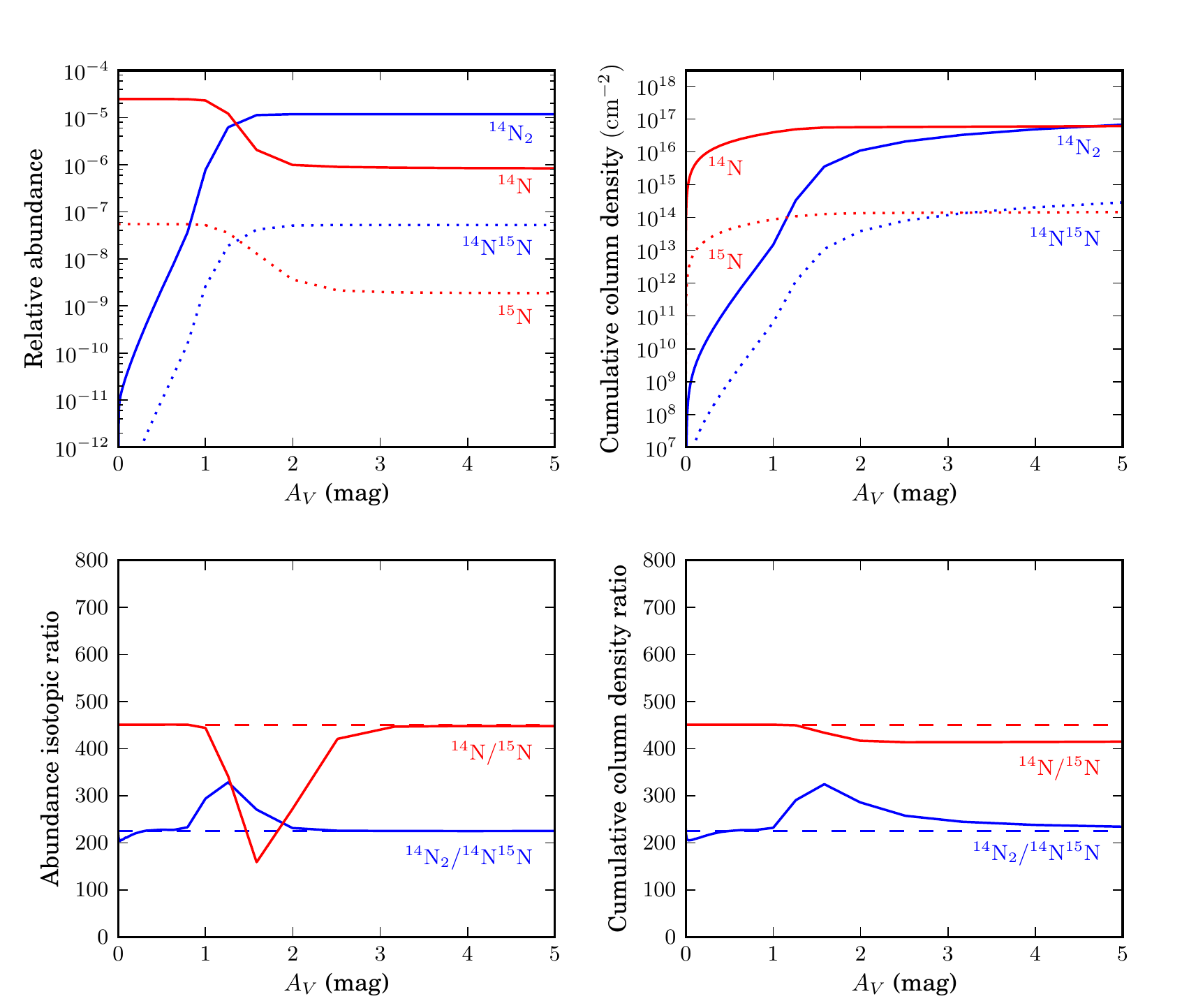}
\caption{\emph{Top:} Abundances relative to H nuclei and left-to-right accumulated column densities of the isotopologues of N and N$_2$ in the modelled interstellar cloud. \emph{Bottom:} Ratios of abundances and column densities of ${}^{14}$N- and ${}^{15}$N-bearing species. Also indicated are the adopted elemental ratios of \ce{^{14}N / ^{15}N} and \ce{^{14}N2 / ^{14}N ^{15}N} (\emph{dashed lines}).}
\label{fig:trans results}
\end{figure*}
Figure~\ref{fig:trans results} shows the calculated abundances of ${}^{14}$N, ${}^{15}$N, \ce{^{14}N2}, and $\mathrm{{}^{14}N{}^{15}N}$ as well as their accumulated column densities, i.e.,
\begin{equation}
  \label{eq:accumulated column density with AV}
  \textrm{Cumulative column density} = \int_0^{z(A_V)} \!\!\!n(\textrm{X})\,dz,
\end{equation}
where $n(\textrm{X})$ is the absolute abundance of species X and $z(A_V)$ is the distance from the clouds edge for a particular extinction.
The trend of these quantities as a function of $A_V$ is qualitatively very similar for either isotopologue. 
Nitrogen is preserved in atomic form near the edge of the cloud due to the intense photodissociating interstellar radiation field.
Increasing shielding with depth into the cloud permits a significant population in molecular form beginning around $A_V=1.5$.

The relative abundances of ${}^{14}$N/${}^{15}$N and \ce{^{14}N2}/$\mathrm{{}^{14}N{}^{15}N}$ calculated by the model are shown in Fig.~\ref{fig:trans results}.
Any fractionation process will alter the isotopic abundances from the assumed elemental ratios ${}^{14}$N/${}^{15}$N=450 and \ce{^{14}N2}/$\mathrm{{}^{14}N{}^{15}N}$=225. 
There is indeed an enhancement of \ce{^{15}N} in atomic form for $A_V$ between 1 and 3\,mag, and its concurrent depletion in the form of \ce{N2}.
This range of extinction corresponds to a column-density of \ce{^{14}N2} sufficiently large to significantly self-shield the ISRF and slow its rate of photodissociation.
The lower column of \ce{^{14}N^{15}N} is not as effectively self-shielding and leads to a larger \ce{^{14}N^{15}N} photodissociation rate and the calculated excess of \ce{^{15}N} atoms.

\change{A modified model was also run neglecting ion-molecule exchange reactions in order to test their contribution to the calculated fractionation.
The modelled abundances were not significantly altered by this modification and the fractionation between $A_V=1$ and 3\,mag is completely dominated by isotope-selective photodissociation due to self-shielding, at least at the adopted temperature of 30\,K.
}
\change{
In contrast, significant enhancements of \ce{^{15}N} in \ce{NH3} and HCN were produced in the models of \citet{charnley2002}, \citet{rodgers2004}, and \citet{wirstrom2012} which simulate dense protostellar cores.
These models included isotope-exchange reactions similar to those adopted here and neglected isotope-selective photodissociation.
They also simulated somewhat colder conditions than ours (10 versus 30\,K), enhancing the rate of the fractionating exchange reactions, and assumed significant depletion of CO relative to \ce{N2} by condensation onto grains, slowing the effect of fractionation-reducing reprocessing to \ce{N2}.
It is then possible that the observed fractionation of N-isotopes may have a combined chemical and photolytic origin.
}

There is no indication in Fig.~\ref{fig:trans results} for fractionation in the abundance of N$_2$ or N in the completely shielded interior of the interstellar cloud model ($A_V\gtrsim 3$).
\change{This is despite the inclusion of isotope-selective ion-molecule exchange reactions} and a 40\% faster rate of cosmic ray induced photodissociation for \ce{^{14}N ^{15}N} relative to \ce{^{14}N2} given the physical parameters selected for this cloud model.
Ultimately, cosmic-ray induced photodissociation is rendered irrelevant by faster and isotopologue-independent destruction mechanisms included in our model:
\begin{equation}
  \label{eq:destruction N2 He+}
  \ce{N2  ->[\ce{He+}] N+ + N}
\end{equation}
and
\begin{equation}
  \label{eq:destruction N2 H3+}
\ce{N2  ->[\ce{H3+}] N2H+ }.
\end{equation}
The combined destruction rate due to these processes exceeds $10^{-13}$\,s$^{-1}$ and is two orders of magnitude faster than cosmic-ray induced photodissociation.

The total-column fractionation of N and \ce{N2} depends on the assumed depth of the modelled cloud, and is also plotted in Fig.~\ref{fig:trans results} as a function of $A_V$.
For our plane-parallel single-side-illuminated model this is maximal for $A_V$ between 1.5 and 2\,mag, but never exceeds a factor of two for atomic or molecular species.

The minor contribution of cosmic-ray induced photodissociation to the nitrogen chemistry in our model does not presuppose a similar conclusion for other molecules, some of which have significantly higher cosmic-ray induced predissociation rates \citep{gredel1989}.

\subsection{Protoplanetary disc model}
\label{sec:protoplanetary disc}

\change{We have run a second chemical model for a vertical slice through a circumstellar disc using the same network of reactions as in Sec.~\ref{sec:interstellar cloud}.
The model setup is identical to that of \citet{visser2009} and \citet{li2013}.
Briefly, we take the disc geometry, dust density and temperature distributions directly from the previous model of \citet{alessio1999} without modification.
This simulates a disc of mass of 0.07\,$M_\odot$ and outer radius 400\,AU surrounding a T~Tauri star of mass 0.5\,$M_\odot$ and radius 2\,$R_\odot$.
The simulated vertical slice is located at a radius of 105\,AU.}

\change{A radiation field with the Draine wavelength dependence is perpendicularly incident on the surface of the slice, 120\,AU above the midplane.
The use of a scaled Draine field to represent the combined UV flux from the central star and the ISRF has been previously investigated in detail for the same disc geometry employed here \citep{van_zadelhoff2003,jonkheid2004}.
From these studies, a Draine field with $\chi=516$ is a satisfactory proxy for a more realistic field including the details of the stellar spectrum at a disc radius of 105\,AU.
Dust grains of radius 1\,$\mu$m are assumed to populate the disc which leads to weaker shielding and a greater penetration of ultraviolet radiation than for the interstellar cloud model of Sec.~\ref{sec:interstellar cloud}.
}

\begin{figure*}
  \centering
  \includegraphics{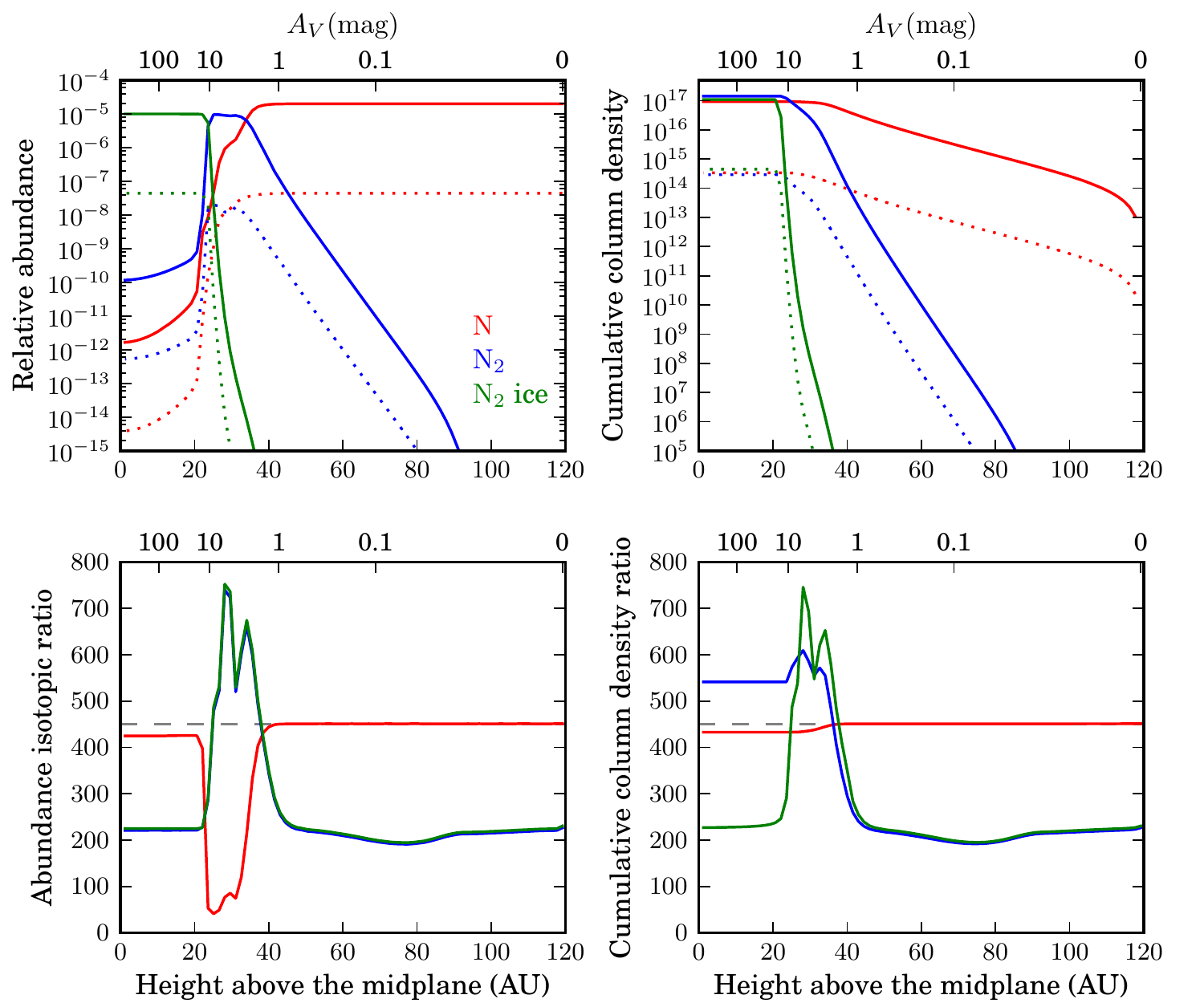}
  \caption{\emph{Top:} Abundances relative to H nuclei and right-to-left accumulated column densities of atomic and molecular nitrogen, and N$_2$ ice in the modelled protoplanetary disc as a function of height above the midplane at a radius of 105\,AU. Also shown for species with a single ${}^{15}$N substitution i.e., ${}^{15}$N and \ce{^{14}N ^{15}N} (\emph{dotted curves}). \emph{Bottom:} Ratios of abundances and column densities of ${}^{14}$N- and ${}^{15}$N-bearing species. In some cases the ice and gaseous \ce{N2} curves coincide. Also indicated is the elemental $\mathrm{{}^{14}N/^{15}N}$ ratio (\emph{dashed line}).}
  \label{fig:disc results N N2}
\end{figure*}

The abundances and cumulative column densities of atomic and molecular nitrogen in the disc model are plotted in Fig.~\ref{fig:disc results N N2}.
The abundance of N$_2$ becomes significant about 40\,AU above the midplane and a rapid freeze-out of gaseous N$_2$ onto dust grains occurs 20\,AU above the midplane, where the temperature falls below 20\,K.
The effect of photodissociative fractionation in the intervening 20 to 40\,AU region is evident in the abundance ratios plotted in Fig.~\ref{fig:disc results N N2}, where the atomic ratio of $\mathrm{{}^{14}N/^{15}N}$ dips by a factor of 10 below the elemental ratio.
However, fractionation of the total atomic N column is negligible because of its low abundance in this region.
As for the interstellar cloud model, cosmic-ray induced photodissociation of N$_2$ is found to be negligible throughout the disc slice, and does not contribute to isotopic fractionation.
\change{Similarly, the contribution of ion-molecule exchange reactions to the calculated chemical abundances was found to be negligible following comparison with an alternative model where these are neglected.}
\change{That is, the ice-phase predominance of nitrogen species in our model for disc heights with sufficiently low-temperature for chemical fractionation to occur suppresses these reactions.}

\begin{figure*}
  \centering
  \includegraphics{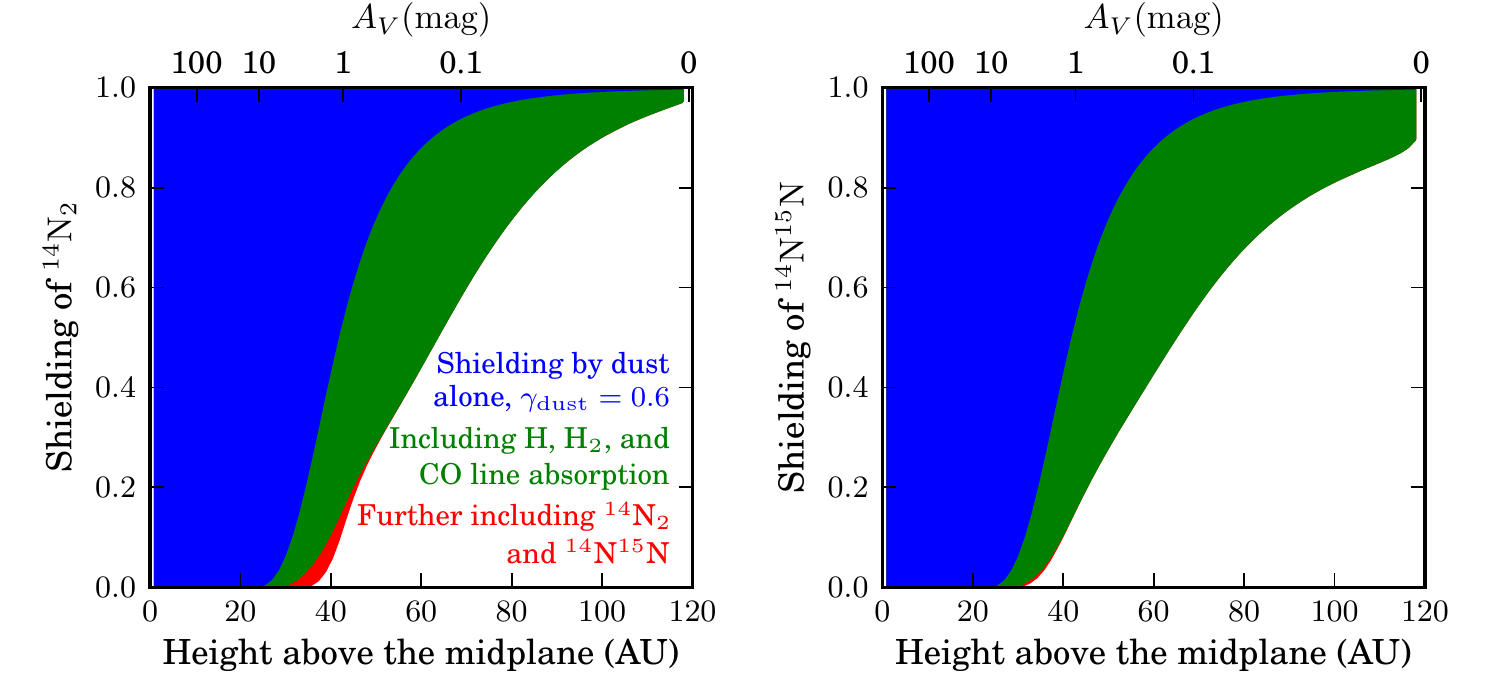}
  \caption{Shielding functions for \ce{^{14}N2} and $\mathrm{{}^{14}N{}^{15}N}$ calculated for the protoplanetary disc model column densities. Partial shielding functions are shown considering dust grains only and with progressively more line absorption.}
  \label{fig:disc model shielding functions}
\end{figure*}

The contribution of different species to the shielding of \ce{^{14}N2} and \ce{^{14}N ^{15}N} photodissociation were calculated as a function of height and $A_V$ and are plotted in Fig.~\ref{fig:disc model shielding functions}.
These calculations include all shielding species in an integral with the form in Eq.~(\ref{eq:def shielding function}) by adopting column densities output from the disc model.
From this, it is evident that shielding by 1\,$\mu$m dust grains (approximated for Fig.~\ref{fig:disc model shielding functions} by assuming $\gamma_\mathrm{dust}=0.6$) and the hydrogen column are comparably effective.
The shielding by CO is quite negligible because of the rarity of its overlaps with N$_2$, evident in Fig.~\ref{fig:cross sections}.
The significance of photodissociation-induced isotopic fractionation of atomic N relies on the relative importance of self-shielding of \ce{^{14}N2} and \ce{^{14}N ^{15}N}.
For \ce{^{14}N2} this becomes important around 40\,AU above the midplane and is comparable to dust and hydrogen shielding between 20 and 35\,AU.
For \ce{^{14}N ^{15}N}, self-shielding never contributes more than 10\% of the total shielding and is about equivalent to its shielding by \ce{^{14}N2}.

\begin{figure*}
  \centering
  \includegraphics{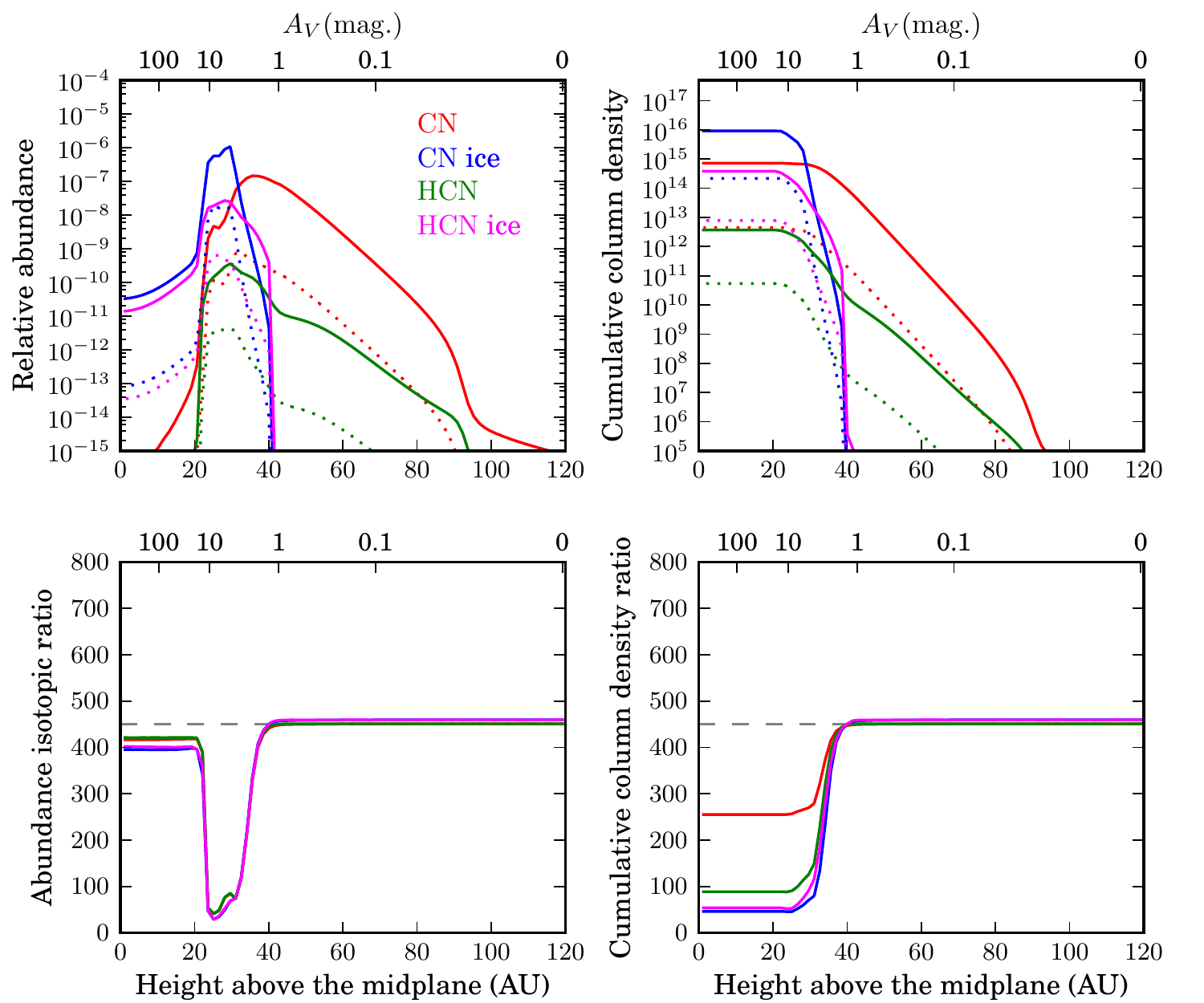}
  \caption{\emph{Top:} Abundances relative to H nuclei and right-to-left accumulated column densities of CN, HCN, and their ices in the modelled protoplanetary disc as a function of height above the midplane at a radius of 105\,AU. Also shown for species with a single ${}^{15}$N substitution (\emph{dotted curves}). \emph{Bottom:} Ratios of abundances and column densities of ${}^{14}$N- and ${}^{15}$N-bearing species. Also indicated is the elemental $\mathrm{{}^{14}N/^{15}N}$ ratio (\emph{dashed line}).}
  \label{fig:disc results CN HCN}
\end{figure*}
The appearance of N and \ce{N2} fractionation between 20 and 40\,AU in Fig.~\ref{fig:disc results N N2} has consequences for other nitrogen bearing species.
For example, Fig.~\ref{fig:disc results CN HCN} shows the abundances, column densities, and isotopic ratios of CN, HCN, and their ices; which are formed primarily from atomic N via the reactions
\begin{equation}
  \label{eq:CN formation}
  \ce{N ->[\ce{CH}] CN},
\end{equation}
and
\begin{equation}
  \label{eq:HCN formation}
  \ce{N ->[\ce{CH_2}] HCN}.
\end{equation}
Additionally, CN and HCN are themselves susceptible to photodissociation and do not survive far from the midplane.
Then, their maximum abundances occur within the region of enhanced ${}^{15}$N atoms and this enhancement is mirrored in their integrated column densities.
The isotopic fractionation of HCN gas and ice columns is plotted in Fig.~\ref{fig:disc results CN HCN} and approaches a factor of 10, whereas the fractionation of gaseous CN is somewhat less because a major portion of its column forms above 40\,AU and outside the region of photo-fractionated N. 

\begin{figure*}
  \centering
  \includegraphics{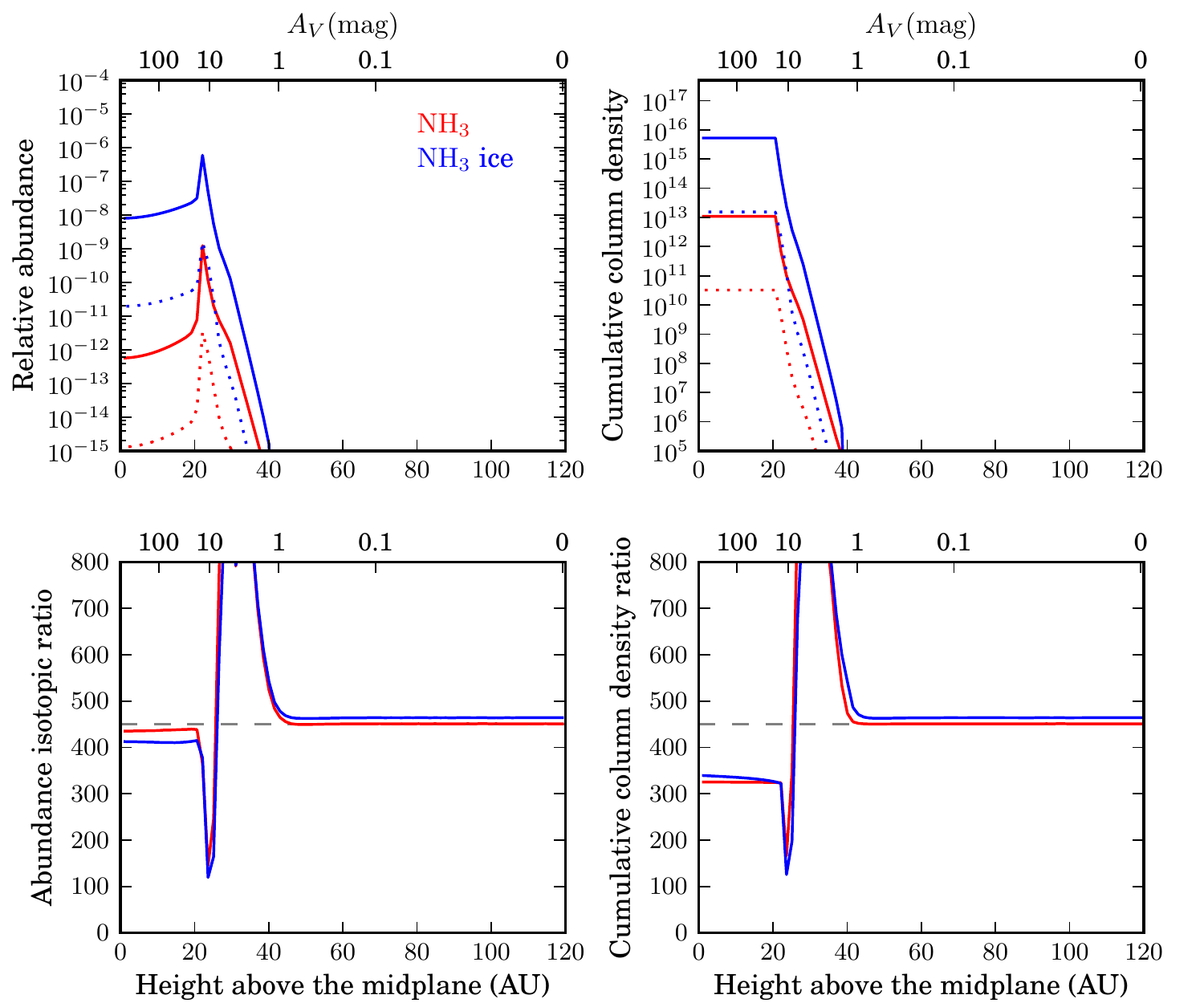}
  \caption{\emph{Top:} Abundances relative to H nuclei and right-to-left accumulated column densities of NH$_3$ and NH$_3$ ice in the modelled protoplanetary disc as a function of height above the midplane at a radius of 105\,AU. Also shown for species with a single ${}^{15}$N substitution (\emph{dotted curves}). \emph{Bottom:} Ratios of abundances and column densities of ${}^{14}$N- and ${}^{15}$N-bearinge species. Also indicated is the elemental $\mathrm{{}^{14}N/^{15}N}$ ratio (\emph{dashed line}).}
  \label{fig:disc results NH3}
\end{figure*}
A different isotope fractionation pattern is evident in Fig.~\ref{fig:disc results NH3} for \ce{NH3}.
The majority of the \ce{NH3} population is frozen onto grain surfaces and its abundance has a sharp maximum at the inner edge of the \ce{^{15}N / ^{14}N} fractionation zone, near 20\,AU.
We do not compute much difference between the integrated column densities of \ce{^{14}NH3} and \ce{^{15}NH3} in either gas or ice phases but the local \ce{^{15}NH3} abundance is significantly enhanced and depleted in the ranges 20--25 and 25--40\,AU, respectively.
The pattern of \ce{NH3} photolytic fractionation is more complex than for CN and HCN due to a more complex formation mechanism.
There are two pathways leading to \ce{NH3} in our model, the first occurring entirely in the gas phase and initiated from molecular \ce{N2}, that is,
\begin{equation}
  \label{eq:NH3 formation path N+}
  \ce{ N2 ->[\ce{He+}] N + N+ ->[n\ce{H2}] NH4+ ->[e^-] NH3 },
\end{equation}
and the second follows from the hydrogenation of atomic N after its condensation onto ice grains,
\begin{equation}
  \label{eq:NH3 formation path ice}
  \ce{N ->[T\lesssim 20\,\textrm{K}] N\ ice ->[n\ce{H}] NH3\ ice }.
\end{equation}
Intermediate species \ce{NH} and \ce{NH2} in the reaction series summarised by Eq.~(\ref{eq:NH3 formation path ice}) are also allowed to evaporate in our model and contribute to the gaseous \ce{NH3} beginning with their ionisation by \ce{H+}, e.g.,
\begin{equation}
  \label{eq:NH3 formation path ice extra}
  \ce{NH2 ->[\ce{H+}] NH2+ ->[\ce{H2}] NH3}.
\end{equation}

The balance between formation routes given in Eqs.~(\ref{eq:NH3 formation path N+}), (\ref{eq:NH3 formation path ice}), and (\ref{eq:NH3 formation path ice extra}) dictates the height-dependent isotopic fractionation of \ce{NH3} in sympathy with the fractionation of their respective precursors, \ce{N2} or N.
The occurrence of formation route (\ref{eq:NH3 formation path N+}) is limited to approximately 20 to 40\,AU above the midplane, where there is an abundance of gaseous \ce{N2}.
The process in Eq.~(\ref{eq:NH3 formation path ice}) requires condensed atomic N and is even more restricted.
That is, sufficiently low temperatures only occur within 25\,AU of the midplane in our model, whereas the necessary source of gaseous N is limited to above 20\,AU.
Then, the \ce{NH3} fractionation pattern resembles that of atomic N between 20 and 25\,AU, and \ce{N2} above this.

Overall, the complexity of photodissociative fractionation of \ce{NH3} deduced here suggests the induced \ce{^{14}NH3 / ^{15}NH3} column density ratio in a particular environment will depend sensitively on its temperature and density profiles.
The photodissociative fractionation of CN and HCN is likely more robust because the formation routes of these molecules are relatively simple.
\change{The enhancement of \ce{^{15}N} in CN and HCN within the self-shielding zone of our model disc ($10\times$) is actually larger than for observations of cometary CN and HCN ($2-3\times$) \citep{jehin2009,mumma2011}.}
This supports the feasibility of photodissociation as a mechanism for cometary fractionation.
The greater fractionation of \ce{^{15}N} and \ce{HCN} relative to \ce{NH3} calculated for our particular protoplanetary disc also suggests that photodissociation may contribute to the explanation of a similar difference observed in the cold prestellar core L1544 \citep{hily-blant2013,bizzocchi2013}, but not necessarily elsewhere in the galaxy \citep{daniel2013}.

\section{Conclusions}
\label{sec:conclusions}

We used a high-resolution theoretical spectroscopic model of \ce{^{14}N ^{15}N} to calculate its dissociation rate in interstellar space due to ultraviolet radiation, as well as the shielding of this radiation by dust grains and gas phase species.
We obtain an unshielded rate of $1.73\times 10^{-10}\,\mathrm{s}^{-1}$ assuming a Draine ISRF and an excitation temperature of 30\,K, which is very near to the rate for \ce{^{14}N2} obtained in our previous study \citep{li2013}.
Comprehensive tabulated shielding functions for \ce{^{14}N2} and \ce{^{14}N ^{15}N} are available online.\footnote{\tt www.strw.leidenuniv.nl/\textasciitilde{}ewine/photo}

To calculate the rate of photodissociation in regions fully shielded from external radiation, the model of \ce{^{14}N2} and \ce{^{14}N ^{15}N} was combined with a detailed theoretical spectrum of H$_2$ emission arising from cosmic rays.
\change{Assuming $\zeta=10^{-16}$\,s$^{-1}$, a variable} rate of $\sim10^{-15}\,\mathrm{s}^{-1}$ was found to depend on the assumed details of the local environment, most significantly, the excitation temperature of N$_2$ and shielding species, the $o\!-\!\mathrm{H}_2\!:\!p\!-\!\mathrm{H}_2$ ratio, and the dust opacity at ultraviolet wavelengths.
Isotope-dependent cosmic-ray induced photodissociation was also investigated, with the most extreme difference occurring when assuming 30\,K, $o\!-\!\mathrm{H}_2\!:\!p\!-\!\mathrm{H}_2=0\!:\!1$, and a low dust absorption cross section appropriate to a protoplanetary disc experiencing grain growth. 
Then, the dissociation rate of \ce{^{14}N ^{15}N} exceeds that of \ce{^{14}N2} by a factor of 8.
This large isotopic difference does not occur at a lower excitation temperature of 10\,K.
We also calculated the rate of CO dissociation by cosmic-ray induced photons previously determined by \citet{gredel1987} using more accurate photodissociation cross sections, and deduced a $\mathord{\sim}40\,\%$ lower value.

Chemical models were run simulating an interstellar cloud and a protoplanetary disc to test the importance of self-shielding of external radiation and cosmic ray initiated photodissociation to the balance of ${}^{14}\mathrm{N}$- and ${}^{15}\mathrm{N}$-bearing species.
An enhancement of atomic ${}^{15}\mathrm{N}$ relative to ${}^{14}\mathrm{N}$ is found at extinctions with $1 \lesssim A_V \lesssim 3$, and is due to more effective self-shielding of the ISRF by \ce{^{14}N2} than by \ce{^{14}N ^{15}N}.
This fractionation is larger for our disc model where weaker dust shielding was assumed to simulate the growth and depletion of small dust grains.
The photodissociation due to cosmic-ray induced ultraviolet radiation is found to be too slow to significantly influence the chemistry in the dark interiors of our models.

The isotopically-dependent photodissociation of N$_2$ in our protoplanetary disc model leads to a significant enhancement in the column densities of \ce{HC^{15}N} relative to \ce{HC^{14}N}.
Our model predicts a value of $\ce{HC^{14}N}/\ce{HC^{15}N}\simeq100$, when assuming an elemental nitrogen abundance ratio of $\ce{{}^{14}N}/\ce{{}^{15}N}=450$.
This corresponds to an isotopic fractionation in $\delta$-notation of ${}^{15}\delta_\ce{HCN}\simeq -800\,\text{\textperthousand}$.
This enhancement mirrors the fractionation of atomic N induced by photodissociation.
Our model predicts a lower fractionation ratio for CN, with $\ce{C^{14}N}/\ce{C^{15}N}\simeq250$.
This is because the region where \ce{CN} forms does not coincide as neatly with the peak of \ce{N2} isotope dependent photodissociation.
The formation chemistry of \ce{NH3} is more complex and can be initiated from both N$_2$ and atomic N.
This species is then less fractionated in our model.

Many assumptions must be made when modelling the chemistry of complex and remote objects such as protoplanetary discs.
However, the calculations underpinning our predictions of photodissociative fractionation are based on reliable laboratory and theoretical spectroscopy and may be resilient to many structural details of astrophysical objects.

\begin{acknowledgements}
  Astrochemistry in Leiden is supported by the Netherlands Research School for Astronomy (NOVA), by a Spinoza grant and grant 648.000.002 from the Netherlands Organisation for Scientific Research (NWO) via the Dutch Astrochemistry Network, and by the European Community's Seventh Framework Programme FP7/2007-2013 under grant agreements 291141 (CHEMPLAN) and 238258 (LASSIE).
  Calculations of the N$_2$ photodissociation cross sections were supported by the Australian Research Council Discovery Program, through Grant Nos. DP0558962 and DP0773050.
  We would also like to thank Simon Bruderer and Xiaohu Li for discussions on the optical properties of protoplanetary dust grains and \ce{^{14}N2} photodissociation, respectively. 
\end{acknowledgements}

\bibliographystyle{aa}
\bibliography{anhrefs}

\begin{thebibliography}{86}
\expandafter\ifx\csname natexlab\endcsname\relax\def\natexlab#1{#1}\fi

\bibitem[{{Adande} \& {Ziurys}(2012)}]{adande2012}
{Adande}, G.~R. \& {Ziurys}, L.~M. 2012, Astrophys. J., 744, 194

\bibitem[{{Aikawa} {et~al.}(2008){Aikawa}, {Wakelam}, {Garrod}, \&
  {Herbst}}]{aikawa2008}
{Aikawa}, Y., {Wakelam}, V., {Garrod}, R.~T., \& {Herbst}, E. 2008, Astrophys.
  J., 674, 984

\bibitem[{Ajello {et~al.}(1989)Ajello, James, Franklin, \&
  Shemansky}]{ajello_etal1989}
Ajello, J.~M., James, G.~K., Franklin, B.~O., \& Shemansky, D.~E. 1989, Phys.
  Rev. A, 40, 3524

\bibitem[{{Al{\'e}on}(2010)}]{aleon2010}
{Al{\'e}on}, J. 2010, Astrophys. J., 722, 1342

\bibitem[{{Bergin} {et~al.}(2002){Bergin}, {Alves}, {Huard}, \&
  {Lada}}]{bergin2002}
{Bergin}, E.~A., {Alves}, J., {Huard}, T., \& {Lada}, C.~J. 2002, Astrophys. J.
  Lett., 570, L101

\bibitem[{{Bizzocchi} {et~al.}(2013){Bizzocchi}, {Caselli}, {Leonardo}, \&
  {Dore}}]{bizzocchi2013}
{Bizzocchi}, L., {Caselli}, P., {Leonardo}, E., \& {Dore}, L. 2013, Astron.
  Astrophys., 555, A109

\bibitem[{{Bohlin} {et~al.}(1978){Bohlin}, {Savage}, \& {Drake}}]{bohlin1978}
{Bohlin}, R.~C., {Savage}, B.~D., \& {Drake}, J.~F. 1978, Astrophys. J., 224,
  132

\bibitem[{{Cecchi-Pestellini} \& {Aiello}(1992)}]{cecchi-pestellini1992b}
{Cecchi-Pestellini}, C. \& {Aiello}, S. 1992, Mon. Not. R. Astron. Soc., 258,
  125

\bibitem[{{Chaparro Molano} \& {Kamp}(2012)}]{chaparro_molano2012a}
{Chaparro Molano}, G. \& {Kamp}, I. 2012, Astron. Astrophys., 537, A138

\bibitem[{{Charnley} \& {Rodgers}(2002)}]{charnley2002}
{Charnley}, S.~B. \& {Rodgers}, S.~D. 2002, Astrophys. J. Lett., 569, L133

\bibitem[{Cleeves {et~al.}(2013)Cleeves, Adams, \& Bergin}]{cleeves2013}
Cleeves, L.~I., Adams, F.~C., \& Bergin, E.~A. 2013, Astrophys. J., 772, 5

\bibitem[{{D'Alessio} {et~al.}(1999){D'Alessio}, {Calvet}, {Hartmann},
  {Lizano}, \& {Cant{\'o}}}]{alessio1999}
{D'Alessio}, P., {Calvet}, N., {Hartmann}, L., {Lizano}, S., \& {Cant{\'o}}, J.
  1999, Astrophys. J., 527, 893

\bibitem[{Dalgarno(2006)}]{dalgarno2006}
Dalgarno, A. 2006, Pro. Natl. Acad. Sci., 103, 12269

\bibitem[{Daniel {et~al.}(2013)Daniel, Gerin, Roueff, Cernicharo, Marcelino,
  Lique, Lis, Teyssier, Biver, \& Bockel'ee-Morvan}]{daniel2013}
Daniel, F., Gerin, M., Roueff, E., {et~al.} 2013, {arXiv}, 1309.5782

\bibitem[{{Draine}(1978)}]{draine1978}
{Draine}, B.~T. 1978, Astrophys. J. Suppl. Ser., 36, 595

\bibitem[{Dressler(1969)}]{dressler1969}
Dressler, K. 1969, Can. J. Phys., 47, 547

\bibitem[{{Floss} {et~al.}(2006){Floss}, {Stadermann}, {Bradley}, {Dai},
  {Bajt}, {Graham}, \& {Lea}}]{floss2006}
{Floss}, C., {Stadermann}, F.~J., {Bradley}, J.~P., {et~al.} 2006, Geochim.
  Cosmochim. Acta, 70, 2371

\bibitem[{{Geers} {et~al.}(2006){Geers}, {Augereau}, {Pontoppidan},
  {Dullemond}, {Visser}, {Kessler-Silacci}, {Evans}, {van Dishoeck}, {Blake},
  {Boogert}, {Brown}, {Lahuis}, \& {Mer{\'{\i}}n}}]{geers2006}
{Geers}, V.~C., {Augereau}, J.-C., {Pontoppidan}, K.~M., {et~al.} 2006, Astron.
  Astrophys., 459, 545

\bibitem[{{Geers} {et~al.}(2007){Geers}, {Pontoppidan}, {van Dishoeck},
  {Dullemond}, {Augereau}, {Mer{\'{\i}}n}, {Oliveira}, \& {Pel}}]{geers2007}
{Geers}, V.~C., {Pontoppidan}, K.~M., {van Dishoeck}, E.~F., {et~al.} 2007,
  Astron. Astrophys., 469, L35

\bibitem[{{Gerin} {et~al.}(2009){Gerin}, {Marcelino}, {Biver}, {Roueff},
  {Coudert}, {Elkeurti}, {Lis}, \& {Bockel{\'e}e-Morvan}}]{gerin2009}
{Gerin}, M., {Marcelino}, N., {Biver}, N., {et~al.} 2009, Astron. Astrophys.,
  498, L9

\bibitem[{Gredel {et~al.}({1987})Gredel, Lepp, \& Dalgarno}]{gredel1987}
Gredel, R., Lepp, S., \& Dalgarno, A. {1987}, Astrophys. J., {323}, L137

\bibitem[{Gredel {et~al.}({1989})Gredel, Lepp, Dalgarno, \&
  Herbst}]{gredel1989}
Gredel, R., Lepp, S., Dalgarno, A., \& Herbst, E. {1989}, Astrophys. J., {347},
  289

\bibitem[{Haverd {et~al.}({2005})Haverd, Lewis, Gibson, \&
  Stark}]{haverd_etal2005}
Haverd, V.~E., Lewis, B.~R., Gibson, S.~T., \& Stark, G. {2005}, J. Chem.
  Phys., {123}, {214304}

\bibitem[{Heays(2011)}]{heays2011}
Heays, A.~N. 2011, PhD thesis, The Australian National University

\bibitem[{Heays {et~al.}(2011)Heays, Dickenson, Salumbides, de~Oliveira,
  Joyeux, Nahon, Lewis, \& Ubachs}]{heays2011b}
Heays, A.~N., Dickenson, G.~D., Salumbides, E.~J., {et~al.} 2011, J. Chem.
  Phys., 135, 244301

\bibitem[{Heays {et~al.}({2009})Heays, Lewis, Stark, Yoshino, Smith, Huber, \&
  Ito}]{heays_etal2009}
Heays, A.~N., Lewis, B.~R., Stark, G., {et~al.} {2009}, J. Chem. Phys., {131},
  194308

\bibitem[{Helm {et~al.}(1993)Helm, Hazell, \& Bjerre}]{helm_etal1993}
Helm, H., Hazell, I., \& Bjerre, N. 1993, Phys. Rev. A, 48, 2762

\bibitem[{{Hezareh} {et~al.}(2008){Hezareh}, {Houde}, {McCoey}, {Vastel}, \&
  {Peng}}]{hezareh2008}
{Hezareh}, T., {Houde}, M., {McCoey}, C., {Vastel}, C., \& {Peng}, R. 2008,
  Astrophys. J., 684, 1221

\bibitem[{{Hily-Blant} {et~al.}(2013){Hily-Blant}, {Bonal}, {Faure}, \&
  {Quirico}}]{hily-blant2013}
{Hily-Blant}, P., {Bonal}, L., {Faure}, A., \& {Quirico}, E. 2013, Icarus, 223,
  582

\bibitem[{{Indriolo} \& {McCall}(2012)}]{indriolo2012}
{Indriolo}, N. \& {McCall}, B.~J. 2012, Astrophys. J., 745, 91

\bibitem[{{Jehin} {et~al.}(2009){Jehin}, {Manfroid}, {Hutsem{\'e}kers},
  {Arpigny}, \& {Zucconi}}]{jehin2009}
{Jehin}, E., {Manfroid}, J., {Hutsem{\'e}kers}, D., {Arpigny}, C., \&
  {Zucconi}, J.-M. 2009, {Earth Moon and Planets}, 105, 167

\bibitem[{Jonkheid {et~al.}({2004})Jonkheid, Faas, van Zadelhoff, \& van
  Dishoeck}]{jonkheid2004}
Jonkheid, B., Faas, F., van Zadelhoff, G., \& van Dishoeck, E. {2004}, Astron.
  Astrophys., {428}, 511

\bibitem[{{Jonkheid} {et~al.}(2006){Jonkheid}, {Kamp}, {Augereau}, \& {van
  Dishoeck}}]{jonkheid2006}
{Jonkheid}, B., {Kamp}, I., {Augereau}, J.-C., \& {van Dishoeck}, E.~F. 2006,
  Astron. Astrophys., 453, 163

\bibitem[{{Le Gal} {et~al.}(2013){Le Gal}, {Hily-Blant}, {Faure}, {Pineau des
  For{\^e}ts}, {Rist}, \& {Maret}}]{le_gal2013}
{Le Gal}, R., {Hily-Blant}, P., {Faure}, A., {et~al.} 2013, {arXiv}, 1311.5313

\bibitem[{{Le Petit} {et~al.}(2006){Le Petit}, Nehmé, {Le Bourlot}, \&
  Roueff}]{le_petit2006}
{Le Petit}, F., Nehmé, C., {Le Bourlot}, J., \& Roueff, E. 2006, Astrophys. J.
  Suppl. Ser., 164, 506

\bibitem[{Lewis {et~al.}(2008{\natexlab{a}})Lewis, Baldwin, Heays, Gibson,
  Sprengers, Ubachs, \& Fujitake}]{lewis_etal2008c}
Lewis, B.~R., Baldwin, K. G.~H., Heays, A.~N., {et~al.} 2008{\natexlab{a}}, J.
  Chem. Phys., 129, 204303

\bibitem[{Lewis {et~al.}(2008{\natexlab{b}})Lewis, Baldwin, Sprengers, Ubachs,
  Stark, \& Yoshino}]{lewis_etal2008a}
Lewis, B.~R., Baldwin, K. G.~H., Sprengers, J.~P., {et~al.} 2008{\natexlab{b}},
  J. Chem. Phys., 129, 164305

\bibitem[{Lewis {et~al.}(2005{\natexlab{a}})Lewis, Gibson, Sprengers, Ubachs,
  Johansson, \& Wahlstr\"om}]{lewis_etal2005b}
Lewis, B.~R., Gibson, S.~T., Sprengers, J.~P., {et~al.} 2005{\natexlab{a}}, J.
  Chem. Phys., 123, 236101

\bibitem[{Lewis {et~al.}(2005{\natexlab{b}})Lewis, Gibson, Zhang,
  Lefebvre-Brion, \& Robbe}]{lewis_etal2005a}
Lewis, B.~R., Gibson, S.~T., Zhang, W., Lefebvre-Brion, H., \& Robbe, J.~M.
  2005{\natexlab{b}}, J. Chem. Phys., 122, 144302

\bibitem[{Lewis {et~al.}(2008{\natexlab{c}})Lewis, Heays, Gibson,
  Lefebvre-Brion, \& Lefebvre}]{lewis_etal2008b}
Lewis, B.~R., Heays, A.~N., Gibson, S.~T., Lefebvre-Brion, H., \& Lefebvre, R.
  2008{\natexlab{c}}, J. Chem. Phys., 129, 164306

\bibitem[{{Li} \& {Lunine}(2003{\natexlab{a}})}]{li2003b}
{Li}, A. \& {Lunine}, J.~I. 2003{\natexlab{a}}, Astrophys. J., 594, 987

\bibitem[{{Li} \& {Lunine}(2003{\natexlab{b}})}]{li2003a}
{Li}, A. \& {Lunine}, J.~I. 2003{\natexlab{b}}, Astrophys. J., 590, 368

\bibitem[{{Li} {et~al.}(2013){Li}, {Heays}, {Visser}, {Ubachs}, {Lewis},
  {Gibson}, \& {van Dishoeck}}]{li2013}
{Li}, X., {Heays}, A.~N., {Visser}, R., {et~al.} 2013, Astron. Astrophys., 555,
  A14

\bibitem[{Liang {et~al.}(2007)Liang, Heays, Lewis, Gibson, \&
  Yung}]{liang_etal2007}
Liang, M.-C., Heays, A.~N., Lewis, B.~R., Gibson, S.~T., \& Yung, Y.~L. 2007,
  Astrophys. J., 664, L115

\bibitem[{Lyons \& Young(2005)}]{lyons_young2005}
Lyons, J. \& Young, E. 2005, Nature, 435, 317

\bibitem[{{Maaskant} {et~al.}(2013){Maaskant}, {Honda}, {Waters}, {Tielens},
  {Dominik}, {Min}, {Verhoeff}, {Meeus}, \& {van den Ancker}}]{maaskant2013}
{Maaskant}, K.~M., {Honda}, M., {Waters}, L.~B.~F.~M., {et~al.} 2013, Astron.
  Astrophys., 555, A64

\bibitem[{{Marty} {et~al.}(2010){Marty}, {Zimmermann}, {Burnard}, {Wieler},
  {Heber}, {Burnett}, {Wiens}, \& {Bochsler}}]{marty2010}
{Marty}, B., {Zimmermann}, L., {Burnard}, P.~G., {et~al.} 2010, Geochim.
  Cosmochim. Acta, 74, 340

\bibitem[{{Mathis} {et~al.}(1977){Mathis}, {Rumpl}, \&
  {Nordsieck}}]{mathis1977}
{Mathis}, J.~S., {Rumpl}, W., \& {Nordsieck}, K.~H. 1977, Astrophys. J., 217,
  425

\bibitem[{{McElroy} {et~al.}(2013){McElroy}, {Walsh}, {Markwick}, {Cordiner},
  {Smith}, \& {Millar}}]{mcelroy2013}
{McElroy}, D., {Walsh}, C., {Markwick}, A.~J., {et~al.} 2013, Astron.
  Astrophys., 550, A36

\bibitem[{{Mumma} \& {Charnley}(2011)}]{mumma2011}
{Mumma}, M.~J. \& {Charnley}, S.~B. 2011, {Annu. Rev. of Astron. Astrophys.},
  49, 471

\bibitem[{Ndome {et~al.}(2008)Ndome, Hochlaf, Lewis, Heays, Gibson, \&
  Lefebvre-Brion}]{ndome_etal2008}
Ndome, H., Hochlaf, M., Lewis, B.~R., {et~al.} 2008, J. Chem. Phys., 129,
  164307

\bibitem[{Niemann {et~al.}({2005})Niemann, Atreya, Bauer, Carignan, Demick,
  Frost, Gautier, Haberman, Harpold, Hunten, Israel, Lunine, Kasprzak, Owen,
  Paulkovich, Raulin, Raaen, \& Way}]{niemann_etal2005}
Niemann, H.~B., Atreya, S.~K., Bauer, S.~J., {et~al.} {2005}, Nature, {438},
  {779}

\bibitem[{{{\"O}berg} {et~al.}(2010){{\"O}berg}, {Qi}, {Fogel}, {Bergin},
  {Andrews}, {Espaillat}, {van Kempen}, {Wilner}, \& {Pascucci}}]{oberg2010}
{{\"O}berg}, K.~I., {Qi}, C., {Fogel}, J.~K.~J., {et~al.} 2010, Astrophys. J.,
  720, 480

\bibitem[{{Padovani} {et~al.}(2009){Padovani}, {Galli}, \&
  {Glassgold}}]{padovani2009}
{Padovani}, M., {Galli}, D., \& {Glassgold}, A.~E. 2009, Astron. Astrophys.,
  501, 619

\bibitem[{{Prasad} \& {Tarafdar}(1983)}]{prasad1983}
{Prasad}, S.~S. \& {Tarafdar}, S.~P. 1983, Astrophys. J., 267, 603

\bibitem[{{Rimmer} {et~al.}(2012){Rimmer}, {Herbst}, {Morata}, \&
  {Roueff}}]{rimmer2012}
{Rimmer}, P.~B., {Herbst}, E., {Morata}, O., \& {Roueff}, E. 2012, Astron.
  Astrophys., 537, A7

\bibitem[{{Roberge} {et~al.}(1981){Roberge}, {Dalgarno}, \&
  {Flannery}}]{roberge1981}
{Roberge}, W.~G., {Dalgarno}, A., \& {Flannery}, B.~P. 1981, Astrophys. J.,
  243, 817

\bibitem[{{Roberge} {et~al.}(1991){Roberge}, {Jones}, {Lepp}, \&
  {Dalgarno}}]{roberge1991}
{Roberge}, W.~G., {Jones}, D., {Lepp}, S., \& {Dalgarno}, A. 1991, Astrophys.
  J. Suppl. Ser., 77, 287

\bibitem[{Rodgers \& Charnley(2004)}]{rodgers2004}
Rodgers, S.~D. \& Charnley, S.~B. 2004, Mon. Not. R. Astron. Soc., 352, 600

\bibitem[{{Rodgers} \& {Charnley}(2008)}]{rodgers2008}
{Rodgers}, S.~D. \& {Charnley}, S.~B. 2008, Astrophys. J., 689, 1448

\bibitem[{R\"ollig {et~al.}({2007})R\"ollig, Abel, Bell, Bensch, Black,
  Ferland, Jonkheid, Kamp, Kaufman, Le~Bourlot, Le~Petit, Meijerink, Morata,
  Ossenkopf, Roueff, Shaw, Spaans, Sternberg, Stutzki, Thi, van Dishoeck, van
  Hoof, Viti, \& Wolfire}]{roellig2007}
R\"ollig, M., Abel, N.~P., Bell, T., {et~al.} {2007}, Astron. Astrophys.,
  {467}, {187+}

\bibitem[{{Savage} {et~al.}(1977){Savage}, {Bohlin}, {Drake}, \&
  {Budich}}]{savage1977}
{Savage}, B.~D., {Bohlin}, R.~C., {Drake}, J.~F., \& {Budich}, W. 1977,
  Astrophys. J., 216, 291

\bibitem[{{Sheffer} {et~al.}(2002){Sheffer}, {Lambert}, \&
  {Federman}}]{sheffer2002}
{Sheffer}, Y., {Lambert}, D.~L., \& {Federman}, S.~R. 2002, Astrophys. J.
  Lett., 574, L171

\bibitem[{{Siebenmorgen} \& {Kr{\"u}gel}(2010)}]{siebenmorgen}
{Siebenmorgen}, R. \& {Kr{\"u}gel}, E. 2010, Astron. Astrophys., 511, A6

\bibitem[{Smith {et~al.}(2009)Smith, Pontoppidan, Young, Morris, \& van
  Dishoeck}]{smith2009}
Smith, R.~L., Pontoppidan, K.~M., Young, E.~D., Morris, M.~R., \& van Dishoeck,
  E.~F. 2009, Astrophys. J., 701, 163

\bibitem[{Spelsberg \& Meyer(2001)}]{spelsberg_meyer2001}
Spelsberg, D. \& Meyer, W. 2001, J. Chem. Phys., 115, 6438

\bibitem[{Sprengers {et~al.}(2005{\natexlab{a}})Sprengers, Reinhold, Ubachs,
  Baldwin, \& Lewis}]{sprengers_etal2005b}
Sprengers, J.~P., Reinhold, E., Ubachs, W., Baldwin, K. G.~H., \& Lewis, B.~R.
  2005{\natexlab{a}}, J. Chem. Phys., 123, 144315

\bibitem[{Sprengers {et~al.}(2005{\natexlab{b}})Sprengers, Ubachs, \&
  Baldwin}]{sprengers_etal2005}
Sprengers, J.~P., Ubachs, W., \& Baldwin, K. G.~H. 2005{\natexlab{b}}, J. Chem.
  Phys., 122, 144301

\bibitem[{Sprengers {et~al.}(2003)Sprengers, Ubachs, Baldwin, Lewis, \&
  Tchang-Brillet}]{sprengers_etal2003}
Sprengers, J.~P., Ubachs, W., Baldwin, K. G.~H., Lewis, B.~R., \&
  Tchang-Brillet, W.-{\"{U}}.~L. 2003, J. Chem. Phys., 119, 3160

\bibitem[{Sprengers {et~al.}(2004)Sprengers, Ubachs, Johansson, L'Huillier,
  Wahlstr\"om, Lang, Lewis, \& Gibson}]{sprengers_etal2004b}
Sprengers, J.~P., Ubachs, W., Johansson, A., {et~al.} 2004, J. Chem. Phys.,
  120, 8973

\bibitem[{Stark {et~al.}(2005)Stark, Huber, Yoshino, Smith, \&
  Ito}]{stark_etal2005}
Stark, G., Huber, K.~P., Yoshino, K., Smith, P.~L., \& Ito, K. 2005, J. Chem.
  Phys., 123, 214303

\bibitem[{Stark {et~al.}(2008)Stark, Lewis, Heays, Yoshino, Smith, \&
  Ito}]{stark_etal2008}
Stark, G., Lewis, B.~R., Heays, A.~N., {et~al.} 2008, J. Chem. Phys., 128,
  114302

\bibitem[{{Sternberg} {et~al.}(1987){Sternberg}, {Dalgarno}, \&
  {Lepp}}]{sternberg1987}
{Sternberg}, A., {Dalgarno}, A., \& {Lepp}, S. 1987, Astrophys. J., 320, 676

\bibitem[{{Terzieva} \& {Herbst}(2000)}]{terzieva2000}
{Terzieva}, R. \& {Herbst}, E. 2000, Mon. Not. R. Astron. Soc., 317, 563

\bibitem[{Tielens({2013})}]{tielens2013}
Tielens, A. G. G.~M. {2013}, Rev. Mod. Phys., {85}, {1021}

\bibitem[{{Tobin} {et~al.}(2012){Tobin}, {Hartmann}, {Bergin}, {Chiang},
  {Looney}, {Chandler}, {Maret}, \& {Heitsch}}]{tobin2012}
{Tobin}, J.~J., {Hartmann}, L., {Bergin}, E., {et~al.} 2012, Astrophys. J.,
  748, 16

\bibitem[{van Dishoeck {et~al.}(2006)van Dishoeck, Jonkheid, \& van
  Hemert}]{van_dishoeck2006}
van Dishoeck, E.~F., Jonkheid, B., \& van Hemert, M.~C. 2006, Faraday Discuss.,
  133, 231

\bibitem[{van Zadelhoff {et~al.}({2003})van Zadelhoff, Aikawa, Hogerheijde, \&
  van Dishoeck}]{van_zadelhoff2003}
van Zadelhoff, G.-J., Aikawa, Y., Hogerheijde, M., \& van Dishoeck, E. {2003},
  Astron. Astrophys., {397}, {789}

\bibitem[{{van Zadelhoff} {et~al.}(2001){van Zadelhoff}, {van Dishoeck}, {Thi},
  \& {Blake}}]{van_zadelhoff2001}
{van Zadelhoff}, G.-J., {van Dishoeck}, E.~F., {Thi}, W.-F., \& {Blake}, G.~A.
  2001, Astron. Astrophys., 377, 566

\bibitem[{Vieitez {et~al.}(2008)Vieitez, Ivanov, de~Lange, Ubachs, Heays,
  Lewis, \& Stark}]{vieitez_etal2008a}
Vieitez, M.~O., Ivanov, T.~I., de~Lange, C.~A., {et~al.} 2008, J. Chem. Phys.,
  128, 134313

\bibitem[{Visser {et~al.}(2011)Visser, Doty, \& van Dishoeck}]{visser2011}
Visser, R., Doty, S.~D., \& van Dishoeck, E.~F. 2011, Astron. Astrophys., 534,
  A132

\bibitem[{Visser {et~al.}(2009)Visser, van Dishoeck, \& Black}]{visser2009}
Visser, R., van Dishoeck, E.~F., \& Black, J.~H. 2009, Astron. Astrophys., 503,
  323

\bibitem[{{Wannier} {et~al.}(1991){Wannier}, {Andersson}, {Olofsson}, {Ukita},
  \& {Young}}]{wannier1991}
{Wannier}, P.~G., {Andersson}, B.-G., {Olofsson}, H., {Ukita}, N., \& {Young},
  K. 1991, Astrophys. J., 380, 593

\bibitem[{{Wilson}(1999)}]{wilson1999}
{Wilson}, T.~L. 1999, {Rep. Prog. Phys.}, 62, 143

\bibitem[{Wirstr\"om {et~al.}(2012)Wirstr\"om, Charnley, Cordiner, \&
  Milam}]{wirstrom2012}
Wirstr\"om, E.~S., Charnley, S.~B., Cordiner, M.~A., \& Milam, S.~N. 2012,
  Astrophys. J. Lett., 757, L11

\bibitem[{Wu {et~al.}(2012)Wu, Judge, Tsai, Lin, Yih, Lo, Fung, Lee, Lewis,
  Heays, \& Gibson}]{wu2011}
Wu, C. Y.~R., Judge, D.~L., Tsai, M.-H., {et~al.} 2012, J. Chem. Phys., 136,
  044301

\end{thebibliography}

\end{document}